\documentclass[twocolumn]{aastex63}

\received{}
\revised{}
\accepted{}

\submitjournal{The Astrophysical Journal}

\usepackage{rotating}
\usepackage{gensymb}

\shorttitle{Electron Preacceleration at Weak ICM Shocks}
\shortauthors{Ha et al.}

\begin{document}

\title{Effects of Multi-scale Plasma Waves on Electron Preacceleration at Weak Quasi-perpendicular Intracluster Shocks}

\author[0000-0001-7670-4897]{Ji-Hoon Ha}
\affil{Department of Physics, School of Natural Sciences UNIST, Ulsan 44919, Korea}
\author[0000-0002-5441-8985]{Sunjung Kim}
\affiliation{Department of Physics, School of Natural Sciences UNIST, Ulsan 44919, Korea}
\author[0000-0002-5455-2957]{Dongsu Ryu}
\affiliation{Department of Physics, School of Natural Sciences UNIST, Ulsan 44919, Korea}
\author[0000-0002-4674-5687]{Hyesung Kang}
\affiliation{Department of Earth Sciences, Pusan National University, Busan 46241, Korea}
\correspondingauthor{Hyesung Kang}
\email{hskang@pusan.ac.kr}

\begin{abstract}
Radio relics associated with merging galaxy clusters indicate the acceleration of relativistic electrons
in merger-driven shocks with low sonic Mach numbers ($M_{\rm s}\lesssim 3$) in the intracluster medium (ICM).
Recent studies have suggested that electron injection to diffusive shock acceleration (DSA) 
could take place through the so-called Fermi-like acceleration in the shock foot of $\beta=P_{\rm gas}/P_{\rm B}\approx 20-100$ shocks
and the stochastic shock drift acceleration (SSDA) in the shock transition of $\beta\approx 1-5$ shocks.
Here, we explore how the SSDA can facilitate electron preacceleration in weak quasi-perpendicular ($Q_{\perp}$) shocks in $\beta\approx 20-100$ plasmas 
by performing particle-in-cell simulations in the two-dimensional domain large enough to properly encompass ion-scale waves.
We find that in supercritical shocks with $M_{\rm s}\gtrsim M_{\rm AIC}^*\sim 2.3$, 
multi-scale waves are excited by the ion and electron temperature anisotropies in the downstream of the shock ramp,
and that through stochastic pitch-angle scattering off the induced waves, electrons are confined in the shock transition for an extended period. 
Gaining energy through the gradient-drift along the motional electric field, electrons
could be preaccelerated all the way to injection to DSA at such ICM shocks.
Our findings imply that the electron DSA process at weak ICM shocks could explain the origin of radio relics.
However, a further investigation of electron acceleration at subcritical shocks with $M_{\rm s}< 2.3$ is called for,
since the Mach numbers of some observed radio relic shocks derived from radio or X-ray observations
are as low as $M_{\rm s}\sim 1.5$.
  
\end{abstract}


\section{Introduction} 
\label{sec:s1}

Shocks with low sonic Mach numbers ($M_{\rm s} \lesssim 5$) are expected to form naturally in the hot intracluster medium (ICM) during the large-scale structure formation of the universe \citep[e.g.,][]{ryu2003, pfrommer2006, skillman2008,vazza2009,hong2014,schaal2015,ha2018a}. 
In particular, weak shocks with $M_{\rm s}\sim 1.5-3$, induced by major mergers of galaxy clusters,
have been detected in X-ray and radio observations \citep[e.g.,][]{markevitch2007, vanweeren2010,bruggen2012,brunetti2014}. 
The so-called radio relics, diffuse elongated radio structures detected in the outskirts of merging clusters,
are interpreted as synchrotron emission from cosmic ray (CR) electrons accelerated 
via diffusive shock acceleration (DSA), also known as the Fermi first-order process, 
in such merger-driven shocks \citep[see e.g.,]
[for a review]{vanweeren2019}.

The acceleration of nonthermal particles at collisionless shocks in tenuous astrophysical plasmas
involves a very broad spectrum of complex kinetic plasma processes \citep[e.g.,][]{drury1983,balogh2013,marcowith2016}. 
It depends on various parameters including the sonic Mach number, $M_{\rm s}$, the plasma beta, $\beta = P_{\rm gas}/P_{\rm B}$ (the ratio of thermal to magnetic pressures), and the obliquity angle,
$\theta_{\rm Bn}$, between the upstream background magnetic field direction and the shock normal.
For weak shocks in the high-$\beta$ ICM under consideration in this study,
it was shown through particle-in-cell (PIC) simulations that
electrons might be preaccelerated and injected to DSA mainly in quasi-perpendicular ($Q_{\perp}$, hereafter) configuration with $\theta_{\rm Bn} \gtrsim 45^{\circ}$ \citep[e.g.,][]{guo2014a,guo2014b}.
On the other hand, at strong shocks in $\beta\sim1$ plasmas such as supernova remnants,
electrons could be injected to DSA and accelerated to high energies even in quasi-parallel ($Q_{\parallel})$ configuration with $\theta_{\rm Bn} \lesssim 45^{\circ}$, owing to the magnetic turbulence excited by CR protons streaming upstream \citep{park2015}.

Two types of electron preacceleration mechanisms operative at weak, high-$\beta$, $Q_{\perp}$-shocks have been explored in the literature so far:
(1) the so-called {\it Fermi-like acceleration} due to the diffusive scattering of electrons between the shock ramp and upstream self-generated waves in the shock foot \citep{matsukiyo2011,guo2014a},
and (2) the so-called {\it stochastic shock drift acceleration} (SSDA) due to the extended gradient-drift of electrons being confined in the shock transition \citep{matsukiyo2015, katou2019,niemiec2019}. 
Figure \ref{fig:f1} illustrates the characteristic structures of $Q_{\perp}$-shocks with the foot and transition zones.

In the Fermi-like acceleration, electrons are energized through
shock drift acceleration (SDA), and reflected by magnetic mirror at the shock ramp \citep{guo2014a}.
Backstreaming electrons self-generate oblique electron-scale waves in {\color{black}the preshock region} 
via the electron firehose instability (EFI) \citep{guo2014b,kang2019,kim2020}.
This is termed as ``Fermi-like'' acceleration, since electrons are scattered between the shock ramp and {\color{black}the self-generated upstream waves}.
It differs from DSA, because electrons do not experience full diffusive transport back and forth 
across the entire shock transition.

\citet[][KRH2019, hereafter]{kang2019} showed that, in high-$\beta$ plasmas, the Fermi-like acceleration may operate only in {\it supercritical} $Q_{\perp}$-shocks with $M_{\rm s} \gtrsim 2.3$, for which the temperature anisotropy due to SDA-reflected electrons is sufficient enough to trigger the EFI. 
In addition, they argued that because of small wavelengths of the EFI-driven waves ($\lambda \lesssim 20 c/\omega_{\rm pe}$, where $c$ is the speed of light and $\omega_{\rm pe}$ is the electron plasma frequency), 
the preacceleration may not proceed all the way to the injection momentum, $p_{\rm inj}\sim 3 p_{\rm i,th}$.
Here, $p_{\rm i, th} = \sqrt{2m_{\rm i}k_{\rm B}T_{\rm i2}}$ is the postshock ion thermal momentum, where $k_{\rm B}$ is the Boltzmann constant and $T_{\rm i2}$ is the postshock ion temperature. 
Hereafter, the subscripts 1 and 2 denote the preshock and postshock quantities, respectively.

Considering that the width of the shock transition zone is several times the gyroradius of 
postshock thermal ions, i.e., $\Delta x_{\rm shock} \sim r_{\rm L,i}(T_{2,i},B_2)$, 
we argue that electron injection to DSA must require multi-scale waves, which can scatter 
and confine electrons with the momentum up to $\sim p_{\rm inj}$ around the shock transition.
Such multi-scale fluctuations could be generated in the shock transition zone 
by the Alfv\'{e}n ion cyclotron (AIC) and ion-mirror instabilities due to the ion temperature anisotropy,
and by the whistler and electron-mirror instabilities due to the electron temperature anisotropy \citep[e.g.,][]{matsukiyo2015,guo2017,katou2019}.
As illustrated in Figure \ref{fig:f1}, the AIC instability induces ion-scale waves propagating parallel to the background magnetic field, 
leading to the shock surface rippling. 
In the SSDA, electrons undergo stochastic pitch-angle scattering off these multi-scale waves and stay confined
for an extended period in the shock transition, leading to much greater energy gain via the gradient-drift
\citep{niemiec2019,trotta2019}.

In the PIC simulations reported by \citet{guo2014a,guo2014b} and KRH2019,
the transverse size of two-dimensional (2D) simulation boxes was $L_y \lesssim r_{\rm L,i}$, 
where $r_{\rm L,i}$ is Larmor radius of incoming ions.
Hence, the simulation domains were not large enough to properly accommodate the emergence 
of ion-scale fluctuations via the AIC instability and the ensuing shock surface rippling.
On the other hand, the 2D PIC simulation by \citet{niemiec2019},
and the 2D and 3D hybrid simulations implemented by test-particle electrons by \citet{trotta2019}
have the domains large enough to include ion-scale fluctuations in the shock transition.
So these authors found that electrons could be preaccelerated all the way to $p_{\rm inj}$
via the SSDA.
In addition, \citet{trotta2019} found that the AIC instability is triggered 
and the ensuing electron preacceleration operates only in supercritical shocks
with the Alfv\'enic Mach number greater than the critical Mach number, $M_{\rm A,crit}\approx 3.5$.
However, note that 
\citet{niemiec2019} considered a shock in a relatively low-$\beta~(\approx 5)$ plasma
because of severe requirements for computational resources,
while \citet{trotta2019} focused on shocks in the solar wind with $\beta\approx 1$.

In this study, we expand the work of KRH2019 by adopting a much larger 2D simulation box in the transverse direction, while keeping basically the same values for other parameters,
for instance, $M_{\rm s} \approx 2 - 3$ and $\theta_{\rm Bn}=53^{\circ}-73^{\circ}$.
To perform the PIC simulations within a reasonable time frame, however,
we choose $\beta=50$ and the ion-to-electron mass ratio, $m_i/m_e=50$, for the fiducial cases
(see Table 1 below).
The main goals of this paper are 
(1) to find the critical Mach number to trigger the AIC instability and the shock surface
rippling in high-$\beta$ shocks, 
and (2) to explore how the SSDA can facilitate the electron preacceleration 
beyond the point where the Fermi-like acceleration saturates.
However, it remains challenging to simulate the true electron injection to DSA beyond $p_{\rm inj}$
in these high-$\beta$ shocks.
The hybrid approach combined with test-particle electron calculations might provide a feasible solution
\citep[e.g.,][]{trotta2019}, 
although kinetic processes on electron scales are not properly emulated in such hybrid simulations.

In a separate paper, \citet{kim2021} (KHRK2021, hereafter), 
we examine the properties of the microinstabilities due the ion and electron temperature anisotropies 
by carrying out a linear stability analysis for wide ranges of
shock parameters. 
In addition, the linear predictions for some models are compared with 2D PIC simulations with periodic boundary conditions. 
Below, we refer to results from that work, when we interpret the properties of plasma waves 
in the shock transition zone.

This paper is organized as follows. 
In Section \ref{sec:s2}, we give a brief overview of the basic physics of $Q_{\perp}$-shocks.
Section \ref{sec:s3} describes the numerical details of PIC simulations, along with the definitions of various parameters. 
In Section \ref{sec:s4}, we present the simulation results, including the shock structure, instability analysis, 
power spectra of self-excited waves, and electron energy spectra.
The dependence of our findings on various model parameters is also discussed in Section \ref{sec:s4}. 
A brief summary follows in Section \ref{sec:s5}.

\section{Basic Physics of $Q_{\perp}$-Shocks}
\label{sec:s2}

\begin{figure}[t]
\vskip -0.0 cm
\hskip -0.0 cm
\centerline{\includegraphics[width=0.48\textwidth]{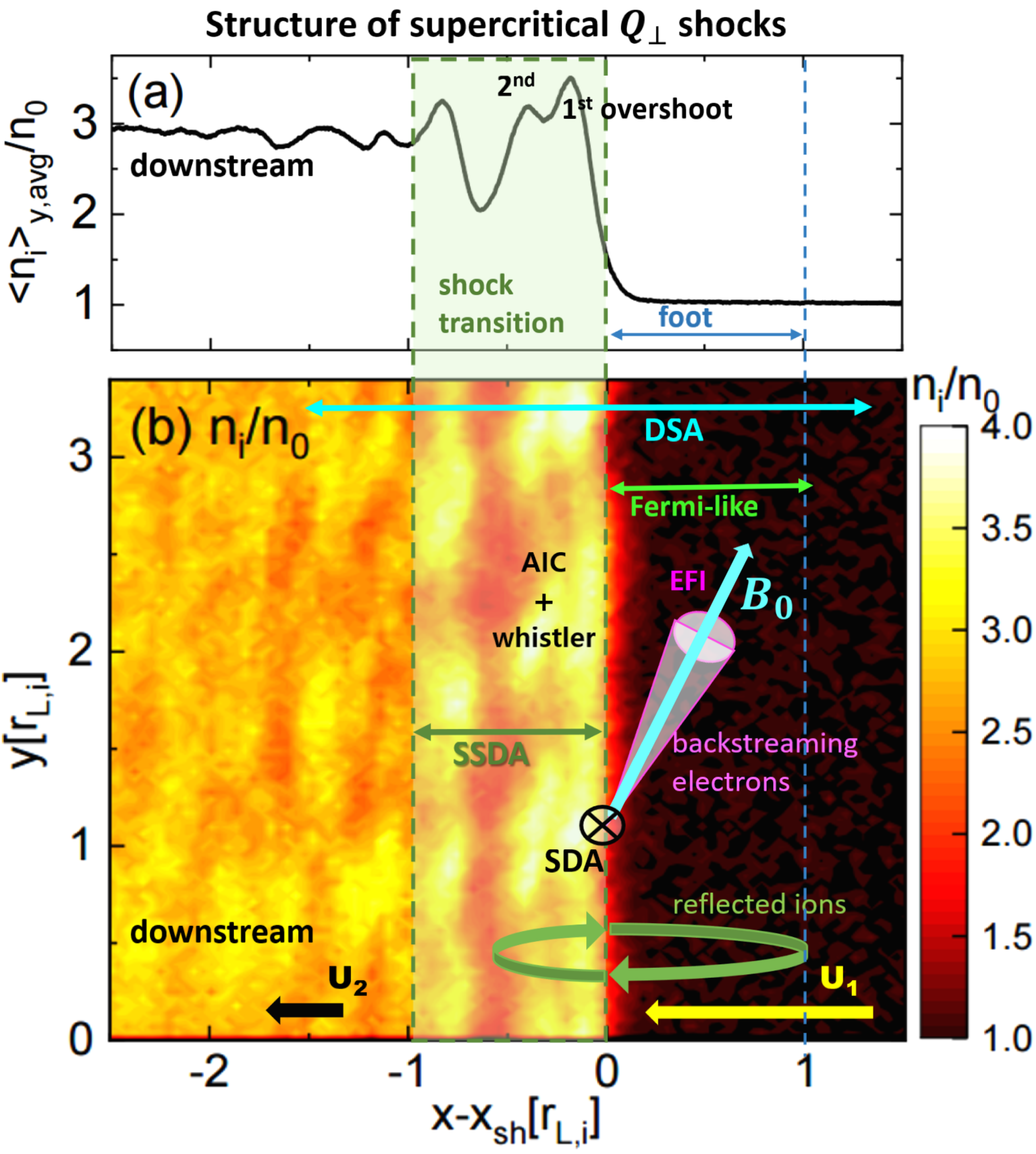}}
\vskip -0.0cm
\caption{(a) Ion number density, $\langle n_{\rm i}\rangle_{y,{\rm avg}}/n_0$, averaged over the $y$-direction, normalized to the upstream ion number density, $n_0$, for a supercritical $Q_{\perp}$-shock (M3.0 model).
(b) Ion number density, $n_{\rm i}(x,y)/n_0$, in the $x-y$ plane for the same model.
The 2D PIC simulation results are shown in the region of $-2.5 \leq (x - x_{\rm sh})/r_{\rm L,i} \leq 1.5$ at $\Omega_{\rm ci}t \sim 32$, where $x_{\rm sh}$ is the shock position.  
The gyromotion of reflected ions (green circular arrows) generates the overshoot/undershoot structure in the shock transition,
while the backstreaming of SDA-reflected electrons (magenta cone) induces the temperature anisotropy and the EFI in {\color{black}the preshock region}.
The colored arrows indicate the regions where DSA (cyan), SSDA (dark green), and Fermi-like acceleration (light green) operate.
The labels for the three instabilities, AIC, whistler, and EFI, are placed in the regions where the respective instabilities are excited.
During a SDA cycle, electrons drift in the negative $z$-direction (into the paper here) anti-parallel to the convection electric field ${\vec E}_{\rm conv}=-(1/c) {\vec U_1}\times {\vec B_0}$.
\label{fig:f1}}
\end{figure}

The physics of kinetic plasma processes in collisionless, $Q_{\perp}$-shocks is complex.
For comprehensive reviews, readers are referred to \citet{balogh2013} and \citet{krasnoselskikh2013}.
An overview of some key processes that are relevant for this study can be found
in KRH2019.

The structures and ensuing excitation of microinstabilities are primarily governed by the dynamics of shock-reflected ions and electrons. 
Figure \ref{fig:f1} illustrates the typical structures of a supercritical $Q_{\perp}$-shock:
(1) the shock foot emerges due to the upstream gyration of reflected ions, and
(2) overshoot/undershoot oscillations develop due to the downstream gyration of those reflected ions in the shock transition zone.
The figure also depicts that the EFI is excited in {\color{black}the preshock region} by SDA-reflected electrons
backstreaming along the background magnetic field \citep[][KRH2019]{guo2014b}, 
whereas the AIC and whistler instabilities are excited along the first and second overshoots, leading
to the shock surface rippling \citep{niemiec2019,trotta2019}.

\subsection{Shock Criticality}
\label{sec:s2.1}

In $Q_{\perp}$-shocks, incoming ions are reflected mainly by the electrostatic potential drop at the shock ramp, 
whereas incoming electrons are reflected by the magnetic mirror force due to converged
magnetic field lines.
In the simplest theory, the shock criticality is directly related to ion reflection at the shock ramp 
when the downstream flow speed normal to the shock exceeds the downstream sound speed, which defines
the condition for the so-called {\it fast first critical Mach number}, $M_{\rm f}^*$ \citep{edmiston1984}.
In addition, there are a few varieties of critical Mach numbers, including the {\it second and third whistler 
critical Mach numbers}, which are related to the emission of dispersive whistler waves and quasi-periodic 
shock-reformation \citep{krasnoselskikh2002,oka2006}.
Obviously, these processes depend on the shock obliquity angle, $\theta_{\rm Bn}$, 
and the plasma beta, $\beta$,
because the ion reflection is affected by anomalous resistivity and microinstabilities in the shock transition.

In \citet{ha2018b} and KRH2019, examining the shock structure, energy spectra of ions and electrons,
and self-excited waves in shock models with $M_{\rm s} \approx 2-3$,
it was suggested that, in high-$\beta~(\sim 100)$ ICM plasmas, 
the first critical Mach number for ion reflection is $M_{\rm s}^*\sim  2.3$ for both $Q_{\parallel}$ and $Q_{\perp}$ shocks,
while the EFI critical Mach number for the excitation of the EFI is also $M_{\rm ef}^*\sim  2.3$ for $Q_{\perp}$-shocks.
These two critical Mach numbers are closely related, since the oscillations in the shock transition due to ion reflection enhance magnetic mirror and electron reflection. 
The critical mach number $M_{\rm s}^*$ is higher than the fast first critical Mach number, $M_{\rm f}^{*} \sim 1$, 
estimated for $\beta \sim 4$ by \citet{edmiston1984} using the fluid description,
in which kinetic processes such as wave excitations and wave-particle interactions were not accounted for. 

Note that $M_{\rm A} \gg M_{\rm s}\approx M_{\rm f}$ in $\beta\gg 1$ plasmas.
In KRH2019, we argued that the critical Mach number for weak ICM shocks should be characterized with $M_{\rm s}$, instead of $M_{\rm A}$,
since primarily the sonic Mach number controls both the shock electrostatic potential drop relevant for ion reflection 
and the magnetic field compression relevant for electron reflection.

In this work, we explore the shock criticality in terms of the shock surface rippling triggered by 
the AIC instability,
using 2D PIC simulations with a transverse dimension large enough to include ion-scale fluctuations.

\subsection{Fermi-like Preacceleration in Shock Foot}
\label{sec:s2.2}

As discussed in the introduction, 
\citet{guo2014a,guo2014b} demonstrated that thermal electrons could be preaccelerated via a 
Fermi-like acceleration in the foot of $M_{\rm s}=3$ shocks in $\beta=20$ ICM plasmas.
The key processes involved in this preacceleration include the following:
(1) a fraction ($\sim 20 \%$) of incoming thermal electrons are reflected and energized through the standard SDA at the shock ramp,
(2) the SDA-reflected electrons backstreaming along the background magnetic field generate oblique 
waves via the EFI due to the electron temperature anisotropy ($T_{e\parallel}/T_{e\perp} > 1$),
and (3) some electrons are scattered back and forth between the shock ramp and 
the EFI-driven upstream waves, gaining energy further through multiple cycles of SDA.
{\color{black}Although the EFI-driven waves can be generated broadly in the upstream region, depending on 
$T_{e\parallel}/T_{e\perp}$, the Fermi-like acceleration occurs primarily within the shock foot.}
In the course of this Fermi-like acceleration, electrons stay in the upstream region of the shock;
hence, they are energized mainly through the gradient-drift along the motional electric field at the shock ramp.
The authors showed that this acceleration is effective for $\beta \gtrsim 20$ shocks,
whereas the EFI is suppressed at low-$\beta$ plasmas due to the strong magnetization of electrons.
Moreover, the EFI is known to be almost independent of the mass ratio, $m_i/m_e$.

Using PIC simulations for $Q_{\perp}$ ICM shocks ($\beta=100$) with different Mach numbers of $M_{\rm s}=2-3$, 
KRH2019 found the followings for shocks with $M_{\rm s}\gtrsim 2.3$.
(1) The fraction of reflected ions increases abruptly and overshoot/undershoot oscillations emerge 
in the shock transition.
(2) The instability parameter for the EFI,
\begin{equation}
\label{eq1}
I_{\rm EFI} \equiv 1 - \frac{T_{\rm e\perp}}{T_{\rm e\parallel}} - \frac{1.27}{\beta_{\rm e\parallel}^{0.95}} > 0,
\end{equation}
increases sharply, where $\beta_{\rm e\parallel} = 8\pi n_{\rm e} k_{\rm B}T_{\rm e\parallel}/B_{0}^2$ is the electron
$\beta$ parallel to the magnetic field.
Also nonpropagating oblique waves appear in {\color{black}the upstream region}.
(3) The upstream electron energy spectrum develops a suprathermal tail for $p\gtrsim 3 p_{\rm th,e}$,
where $p_{\rm th,e}$ is the postshock thermal electron momentum.
These features all imply that this Fermi-like acceleration could be effective only for supercritical 
$Q_{\perp}$-shocks with $M_{\rm s}\gtrsim M_{\rm ef}^*\sim 2.3$.

Considering that the Mach numbers of some observed radio relics, derived from radio or X-ray observations,
seem to indicate CR electron acceleration at subcritical shocks with
$M_{\rm s} \sim 1.5-2.3$ \citep[e.g.,][]{vanweeren2019}, 
KRH2019 argued that a full understanding for the origin of radio relics has yet to come.

Furthermore, analyzing self-excited waves in the shock foot, KRH2019 found that nonpropagating oblique waves 
with $\lambda \sim 15-20 c/ \omega_{\rm pe}$ are dominantly excited, 
and that the scattering of electrons by those waves reduces the temperature anisotropy and stabilize the EFI.
Thus, we suggested that the preacceleration of electrons by the Fermi-like
acceleration involving multiple cycles of SDA may not proceed all the way to DSA in weak ICM shocks.
This calls for additional mechanisms that could energize electrons beyond the point where
the Fermi-like acceleration ceases to operate.

\subsection{Stochastic Shock Drift Acceleration in Shock Transition}
\label{sec:s2.3}

In the transition zone of supercritical $Q_{\perp}$-shocks, three kinds of microinstabilities 
could be excited \citep[e.g.][]{guo2017,katou2019}: 
(1) the AIC instability due to the ion temperature anisotropy ($\mathcal{A}_i\equiv T_{\rm i\perp}/T_{\rm i\parallel} > 1$) induced by the shock-reflected ions that are advected downstream;
(2) the whistler instability due to the electron temperature anisotropy ($\mathcal{A}_e\equiv T_{\rm e\perp}/T_{\rm e\parallel} > 1$) induced by the magnetic field compression at the shock ramp; 
(3) the mirror instabilities due to $\mathcal{A}_i > 1$ and/or $\mathcal{A}_e > 1$. 
The AIC and whistler instabilities excite waves propagating predominantly in the direction parallel to the background magnetic field, 
while the mirror instabilities induce nonpropagating oblique waves.

\citet{katou2019} proposed that the electron preacceleration via SDA could be extended by stochastic 
pitch-angle scattering off these multi-scale waves, because electrons are trapped much longer 
in the shock transition zone.
They coined the term ``stochastic shock drift acceleration (SSDA)'' for such acceleration.
Then, \citet{niemiec2019} performed a 2D PIC simulation for $M_{\rm s}=3$ shock with $\beta=5$,
$\theta_{\rm Bn}=75^{\circ}$ and $m_i/m_e=100$.
They observed the emergence of shock surface rippling, accompanied by the plasma waves driven by the three
kinds of instabilities in the shock transition, as well as the EFI-driven obliques waves in {\color{black}the preshock region}.
They also saw the development of the suprathermal tails in both the upstream and downstream energy 
spectra of electrons that extend slightly beyond $p_{\rm inj}$ by the end of their simulation.

Furthermore, \citet{trotta2019} performed 2D and 3D hybrid simulations with test-particle electrons for $Q_{\perp}$-shocks with $M_{\rm s}=2.9-6.6$, $\beta\approx 1$, and $\theta_{\rm Bn}=80^{\circ}-87^{\circ}$. 
In typical hybrid simulations, ions are treated kinetically, while electrons are treated as a charge-neutralizing fluid. So this type of simulations cannot properly capture electron-driven instabilities.
With that caveat, they observed 
that shock surface fluctuations develop on ion scales, and that test-particles electrons could be preaccelerated well 
beyond $p_{\rm inj}$ at supercritical shocks with the Alfv\'enic Mach number greater than the critical Mach number, $M_{\rm A,crit}\approx 3.5$.
Note that they considered $\beta\approx 1$ shocks, so $M_{\rm s,crit}\approx M_{\rm A,crit}$.

\section{Numerical Setup for PIC Simulations}
\label{sec:s3}

\begin{deluxetable*}{ccccccccccccc}[t]
\tablecaption{Model Parameters for PIC Simulations \label{tab:t1}}
\tabletypesize{\scriptsize}
\tablecolumns{13}
\tablenum{1}
\tablewidth{0pt}
\tablehead{
\colhead{Model Name} &
\colhead{$M_{\rm s}$} &
\colhead{$M_{\rm A}$} &
\colhead{$u_0/c$} &
\colhead{$\theta_{\rm Bn}$} &
\colhead{$\beta$} &
\colhead{$T_{\rm e0} = T_{\rm i0} [\rm K(keV)]$} &
\colhead{$m_i/m_e$} &
\colhead{$L_x [c/\omega_{\rm pe}]$} &
\colhead{$L_y [c/\omega_{\rm pe}]$} &
\colhead{$\Delta x [c/\omega_{\rm pe}]$} &
\colhead{$t_{\rm end} [\omega_{\rm pe}^{-1}]$}&
\colhead{$t_{\rm end} [\Omega_{\rm ci}^{-1}]$}
}
\startdata
M2.0 & 2.0 & 12.9 & 0.038 & $63^{\circ}$ &50 & $10^8(8.6)$ & 50 & $3200$ & 310 & 0.1 & $4.2\times 10^4$ & 32\\
M2.15 & 2.15 & 13.9 & 0.042 & $63^{\circ}$ &50 & $10^8(8.6)$ & 50 & $3200$ & 310 & 0.1 & $4.2\times 10^4$ & 32\\
M2.3 & 2.3 & 14.8 & 0.046 & $63^{\circ}$ &50 & $10^8(8.6)$ & 50 & $3200$ & 310 & 0.1 & $4.2\times 10^4$ & 32\\
M2.5 & 2.5 & 16.1 & 0.053 & $63^{\circ}$ &50 & $10^8(8.6)$ & 50 & $3200$ & 310 & 0.1 & $4.2\times 10^4$ & 32\\
M2.8 & 2.8 & 18.1 & 0.061 & $63^{\circ}$ &50 & $10^8(8.6)$ & 50 & $3200$ & 310 & 0.1 & $4.2\times 10^4$ & 32\\
M3.0 & 3.0 & 19.4 & 0.068 & $63^{\circ}$ &50 & $10^8(8.6)$ & 50 & $3200$ & 310 & 0.1 & $4.2\times 10^4$ & 32\\
\hline
M2.0-m100 & 2.0 & 12.9 & 0.027 & $63^{\circ}$ &50 & $10^8(8.6)$ & 100 & $2000$ & 440 & 0.1 & $4.2\times 10^4$ & 20\\
M2.3-m100 & 2.3 & 14.8 & 0.0325 & $63^{\circ}$ &50 & $10^8(8.6)$ & 100 & $2000$ & 440 & 0.1 & $4.2\times 10^4$ & 20\\
M3.0-m100 & 3.0 & 19.4 & 0.047 & $63^{\circ}$ &50 & $10^8(8.6)$ & 100 & $2000$ & 440 & 0.1 & $4.2\times 10^4$ & 20\\
\hline
M2.0-$\beta$20 &  2.0 & 8.2 & 0.038  & $63^{\circ}$ & 20 & $10^8(8.6)$ & 50 & $3200$ & 200 & 0.1 & $2.6\times 10^4$ & 32\\
M2.3-$\beta$20 &  2.3 & 9.4 & 0.046 & $63^{\circ}$ & 20 & $10^8(8.6)$ & 50 & $3200$ & 200 & 0.1 & $2.6\times 10^4$ & 32\\
M3.0-$\beta$20 & 3.0 & 12.3 & 0.068 & $63^{\circ}$ & 20 & $10^8(8.6)$ & 50 & $3200$ & 200 & 0.1 & $2.6\times 10^4$ & 32\\
M2.0-$\beta$100 &  2.0 & 18.2 & 0.038 & $63^{\circ}$ & 100 & $10^8(8.6)$ & 50 & $2000$ & 440 & 0.1 & $3.0\times 10^4$ & 20\\
M2.3-$\beta$100 &  2.3 & 21.0 & 0.046 & $63^{\circ}$ & 100 & $10^8(8.6)$ & 50 & $2000$ & 440 & 0.1 & $3.0\times 10^4$ & 20\\
M3.0-$\beta$100 & 3.0 & 27.4 & 0.068 & $63^{\circ}$ & 100 & $10^8(8.6)$ & 50 & $2000$ & 440 & 0.1 & $3.0\times 10^4$ & 20\\
\hline
M2.0-$\theta$53 & 2.0 & 12.9 & 0.038 & $53^{\circ}$ &50 & $10^8(8.6)$ & 50 & $3200$ & 310 & 0.1 & $4.2\times 10^4$ & 32\\
M2.0-$\theta$73 & 2.0 & 12.9 & 0.038 & $73^{\circ}$ &50 & $10^8(8.6)$ & 50 & $3200$ & 310 & 0.1 & $4.2\times 10^4$ & 32\\
M2.3-$\theta$53 & 2.3 & 14.8 & 0.046 & $53^{\circ}$ &50 & $10^8(8.6)$ & 50 & $3200$ & 310 & 0.1 & $4.2\times 10^4$ & 32\\
M2.3-$\theta$73 & 2.3 & 14.8 & 0.046 & $73^{\circ}$ &50 & $10^8(8.6)$ & 50 & $3200$ & 310 & 0.1 & $4.2\times 10^4$ & 32\\
M3.0-$\theta$53 & 3.0 & 19.4 & 0.068 & $53^{\circ}$ &50 & $10^8(8.6)$ & 50 & $3200$ & 310 & 0.1 & $4.2\times 10^4$ & 32\\
M3.0-$\theta$73 & 3.0 & 19.4 & 0.068 & $73^{\circ}$ &50 & $10^8(8.6)$ & 50 & $4000$ & 310 & 0.1 & $6.6\times 10^4$ & 50\\
\hline
\enddata
\vspace{-0.8cm}
\end{deluxetable*}

The numerical code and setup for PIC simulations are the same as those adopted in KRH2019,
except that here the 2D simulation domain in unit of $r_{\rm L,i}$ is about 8 times larger
in the transverse direction.
TRISTAN-MP code in 2D planar geometry is used \citep[][]{buneman1993,spitkovsky2005}.
An ion-electron plasma with Maxwell distributions moves with the bulk velocity ${\bf{u_0}} = - u_0 \mathbf{\hat{x}}$ toward a reflecting wall at the leftmost boundary ($x = 0$), 
so a shock propagates toward the $+\mathbf{\hat{x}}$ direction. 
A uniform background magnetic field, ${\bf{B_0}}$, lies in the $x$-$y$ plane (shock coplanarity
plane), and the angle between ${\bf{B_0}}$ and the shock normal is the shock obliquity angle, $\theta_{\rm Bn}$.
When ${\bf{B_0}}$ is perpendicular to the simulation plane, the parallel propagating modes such as AIC waves 
are known to be suppressed \citep{burgess2007}. So we focus on the `in-plane' magnetic field configuration in this study.
In addition, to insure the zero initial electric field in the flow frame, the initial electric field is set as $\bf{E_0} = - \bf{u_0}/c \times \bf{B_0}$ along the $+\bf{\hat{z}}$ direction in the simulation frame. 

The incoming plasma is specified by the following parameters relevant for the ICM: 
$n_{\rm i0}=n_{\rm e0}=10^{-4}~{\rm cm}^{-3}$, 
$k_{\rm B}T_{\rm i0} = k_{\rm B}T_{\rm e0} = 8.6$~keV, and $\beta=20-100$.
The magnetic field strength of the incoming flow is given by
$B_0= \sqrt{8\pi k_B (n_{\rm i0}T_{\rm i0}+n_{\rm e0}T_{\rm e0})/\beta}$.
Reduced ion-to-electron mass ratios, $m_i/m_e = 50-100$, are used due to the limitation of available computational resources 
(where $m_ec^2=0.511$~MeV).
Then, the sound speed of the incoming plasma is defined as $c_{\rm s0}= \sqrt{2\Gamma k_{B}T_{\rm i0}/m_{i}}$
(where $\Gamma = 5/3$ is the adiabatic index),
while the Alfv\'en speed becomes $v_{\rm A0}\approx B_0/\sqrt{4\pi n_{\rm i0} m_i}$ in the upstream region.

In weakly magnetized plasmas, the sonic Mach number of the shock induced in such setup
can be estimated as
\begin{equation}
\label{eq:e2}
M_{\rm s} \equiv \frac{u_{\rm sh}}{c_{\rm s0}} \approx \frac{u_0}{c_{\rm s0}} \frac{r}{r-1},
\end{equation}
where $r= (\Gamma + 1)/(\Gamma - 1 + 2/M^2_{\rm s})$ is the Rankine–Hugoniot compression ratio across the shock.
Then, the Alfv\'en Mach number of the shock can be calculated as
$M_{\rm A} \equiv {u_{\rm sh}}/{v_{\rm A0}} \approx \sqrt{\Gamma \beta /2}\cdot M_{\rm s}$.
In addition, the fast Mach number is defined as 
$M_{\rm f} \equiv {u_{\rm sh}}/{v_{\rm f0}}$,
where $v_{\rm f0}^2= \{ (c_{\rm s0}^2 + v_{\rm A0}^2) + [ (c_{\rm s0}^2 + v_{\rm A0}^2)^2 - 4 c_{\rm s0}^2 v_{\rm A0}^2 \cos^2 \theta_{\rm Bn}]^{1/2} \}/2$.
In the limit of high $\beta$ (i.e., $c_{\rm s0}\gg v_{\rm A0}$), $M_{\rm f}\approx M_{\rm s}$.

The parameters of our shock models are given in Table \ref{tab:t1}.
Models with different $M_{\rm s}$ are named with the combination of the letter ``M'' and sonic
Mach numbers (e.g., the M3.0 model has $M_{\rm s}=3.0$).
Six models, M2.0-M3.0, in the top group represent our `fiducial' models with
$\beta=50$, $\theta_{\rm Bn}=63^{\circ}$, and $m_i/m_e=50$.
Models with the parameters different from the fiducial values have the names that are appended by a character for the specific parameter and its value.
For example, the M3.0-m100 model has $m_i/m_e=100$, while the M3.0-$\theta$53 model has $\theta_{\rm Bn}=53^{\circ}$.

In PIC simulations, kinetic plasma processes for different species are followed on different length scales, 
the electron skin depth, $c/\omega_{\rm pe}$, and the ion skin depth, $c/\omega_{\rm pi}$.
Here, $\omega_{\rm pe} = \sqrt{4\pi e^{2}n_e/m_e}$ and $\omega_{\rm pi} = \sqrt{4\pi e^{2}n_i/m_i}$ are the electron and ion plasma frequencies, respectively.  
On the other hand, the shock structure, which is governed by the ion dynamics, 
evolves in the timescales of the ion gyration period, $\Omega_{\rm ci}^{-1} = m_i c/eB_0\propto m_i \sqrt{\beta}$,
and varies on the length scales of the Larmor radius of incoming ions, 
$r_{\rm L,i} \equiv {u_0}/{\Omega_{\rm ci}} \approx 91 (c/\omega_{\rm pe})\cdot (M_{\rm s}/3)
\sqrt{\beta/50}\sqrt{(m_i/m_e)/50}$.

The simulation results are presented mainly in units of the electron skin depth, $c/\omega_{\rm pe}$,
defined with the incoming electron density $n_{e0}$. 
In addition, the ion scales, $c/\omega_{\rm pi}$, $\Omega_{\rm ci}$ and $r_{\rm L,i}$, defined with 
$n_{\rm i0}$, $B_0$ and $u_0$, are also used when appropriate.

Table \ref{tab:t1} lists the size of the 2D simulation domain, $L_{x}$ and  $L_{y}$, 
in the $9^{th}$ and $10^{th}$ columns, respectively.
In the M3.0 model, for example, the transverse dimension is
$L_{y}=310 c/\omega_{\rm pe} \approx 3.4 r_{\rm L,i}$.
All the simulations have the spatial resolution of $\Delta x = \Delta y = 0.1 c/\omega_{\rm pe}$
and include 32 particles (16 per species) per cell.
{\color{black}
We point that \citet{guo2014b} considered $M_{\rm s}=3$ models with $\beta=20-200$ and $k_B T_{\rm e0}=86$~keV (10 times hotter than here),
and showed that the results of their PIC simulations with 32 particles per cell were similar to those with 64 particles per cell. Hence, we expect that the overall results of our PIC simulations are reasonably converged. 
We note that \citet{niemiec2019} performed their simulation with 40 particles per cell for a $M_{\rm s}=3$ model with $\beta=5$ and $k_B T_{\rm e0}=43$~keV (5 times hotter than here).}
The end of simulation time, $t_{\rm end}$, is given in the $12^{th}$ and $13^{th}$ columns.
The time step is $\Delta t = 0.045 \omega_{\rm pe}^{-1}$.

\section{Results}
\label{sec:s4}

\begin{figure}[t]
\vskip -0.2 cm
\hskip -0.2 cm
\centerline{\includegraphics[width=0.52\textwidth]{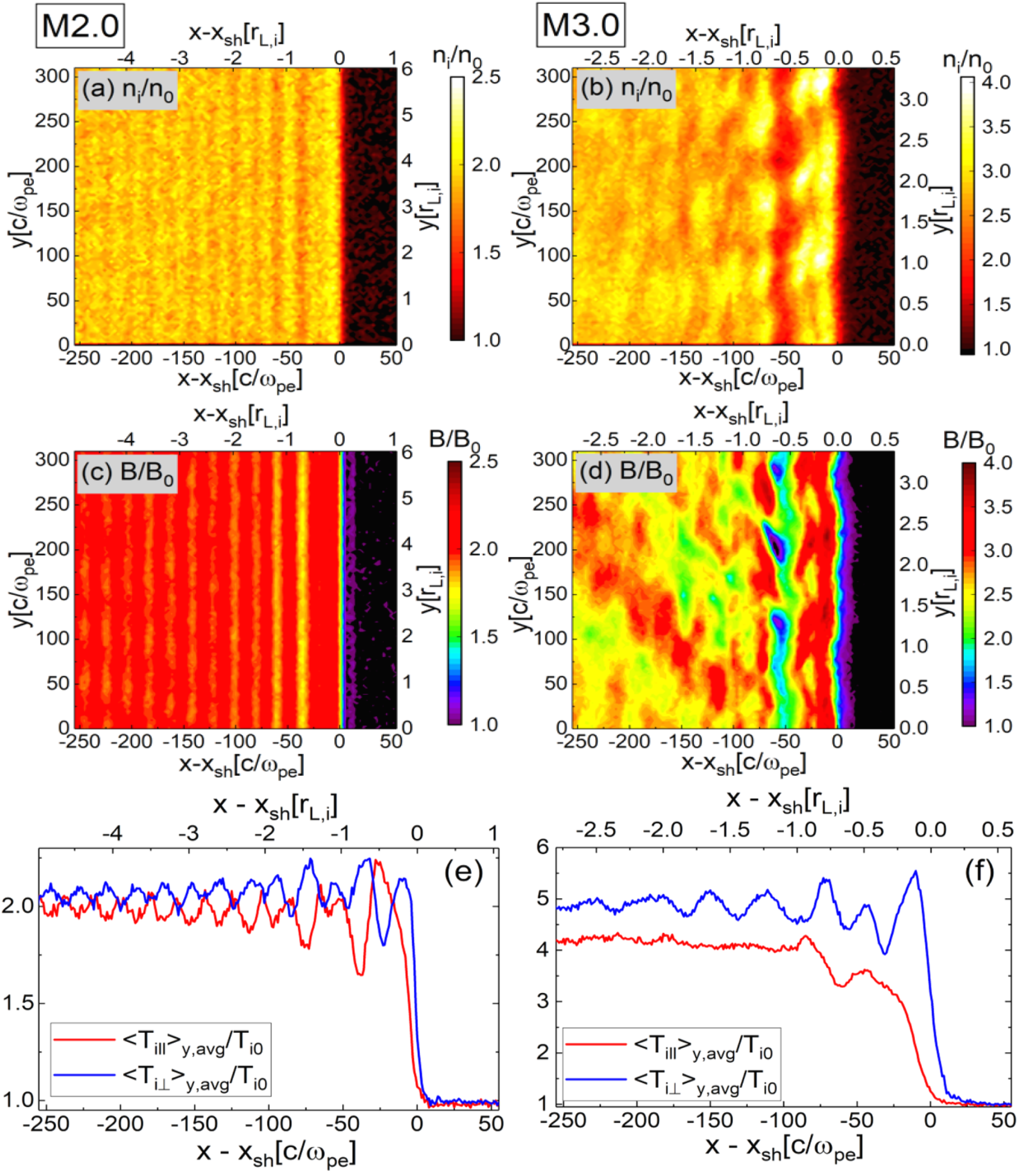}}
\vskip -0.0cm
\caption{Shock structure in the M2.0 and M3.0 models in the region of $-255 \leq (x - x_{\rm sh})\omega_{\rm pe}/c \leq 55$ at $\Omega_{\rm ci}t \sim 32$. 
Panels (a)$-$(b) show the ion number density, $n_{\rm i}/n_0$. 
Panels (c)$-$(d) show the magnetic field strength, $B/B_0$. 
Panels (e)$-$(f) show the ion temperature, 
$\langle T_{\rm i\parallel}\rangle_{y,{\rm avg}}/T_{\rm i0}$ (red), and 
$\langle T_{\rm i\perp}\rangle_{y,{\rm avg}}/T_{\rm i0}$ (blue), averaged over the $y$-direction. 
{\color{black}Here $r_{\rm L,i} \approx 91 (c/\omega_{\rm pe})\cdot (M_{\rm s}/3)$.}
\label{fig:f2}}
\end{figure}

\subsection{Shock Criticality and Surface Rippling}
\label{sec:s4.1}

To understand the shock structure, we first look at the 2D spatial distributions of the ion number density
and the magnetic field strength for the M2.0 (subcritical) and M3.0 (supercritical) models 
in Figure \ref{fig:f2}(a)$-$(d).
In the M3.0 model, overshoot/undershoot oscillations develop and ripples appear in the shock transition zone along the shock surface, 
while the overall shock structure is relatively smooth in the M2.0 model. 
Panels (e) and (f) of Figure \ref{fig:f2} show that the ion temperature anisotropy, $\mathcal{A}_i > 1$, is generated in the shock transition due to the shock-reflected ions in the M3.0 model,
while $\mathcal{A}_i \approx 1$ in the M2.0 model.

Based on the findings of KHRK2021 and previous studies \citep[e.g.,][and references therein]{lowe2003, matsukiyo2015, niemiec2019, trotta2019}, we interpret that the ripples along the shock surface are induced
by the AIC instability.  
{\color{black}In the fiducial M3.0 model, the characteristic length of the ripples is 
$\lambda_{\rm ripple} \sim 14 c/\omega_{\rm pi}\sim 1.1 r_{\rm L,i}\sim 0.8\lambda_{\rm AIC}$, 
where $\lambda_{\rm AIC} \sim 18 c/\omega_{\rm pi}\sim 1.4 r_{\rm L,i}$ is the wavelength of the AIC-driven waves with the maximum growth rate.}
{\color{black}The simulation domain is periodic in the $y$-direction and
the transverse dimension of all the $M_{\rm s}=3.0$ models in Table \ref{tab:t1} 
is $L_{\rm y}\approx 3.4 r_{\rm L,i}\sim 2.4-2.6 \lambda_{\rm AIC}$.
Hence, on average about $2-3$ waves are expected to develop along the $y$-direction (see also Figure \ref{fig:f7}(g)$-$(i)), resulting in $\lambda_{\rm ripple} \sim 0.8-1.3\lambda_{\rm AIC}$.}
In addition, the rippling waves propagate along the shock surface with the Alfv\'{e}n speed in the shock overshoot, $v_{\rm A, os} = B_{\rm os}/\sqrt{4\pi n_{\rm os} m_{\rm i}} \sim 0.009 c$, where $B_{\rm os}$ and $n_{\rm os}$ are the magnetic field strength and the ion number density of the shock overshoot, respectively. 
Hence, we regard that the rippling waves have the characteristics of the waves driven by the AIC instability \citep[][]{lowe2003}. 

\begin{figure}[t]
\vskip -1.0 cm
\hskip -0.2 cm
\centerline{\includegraphics[width=0.35\textwidth]{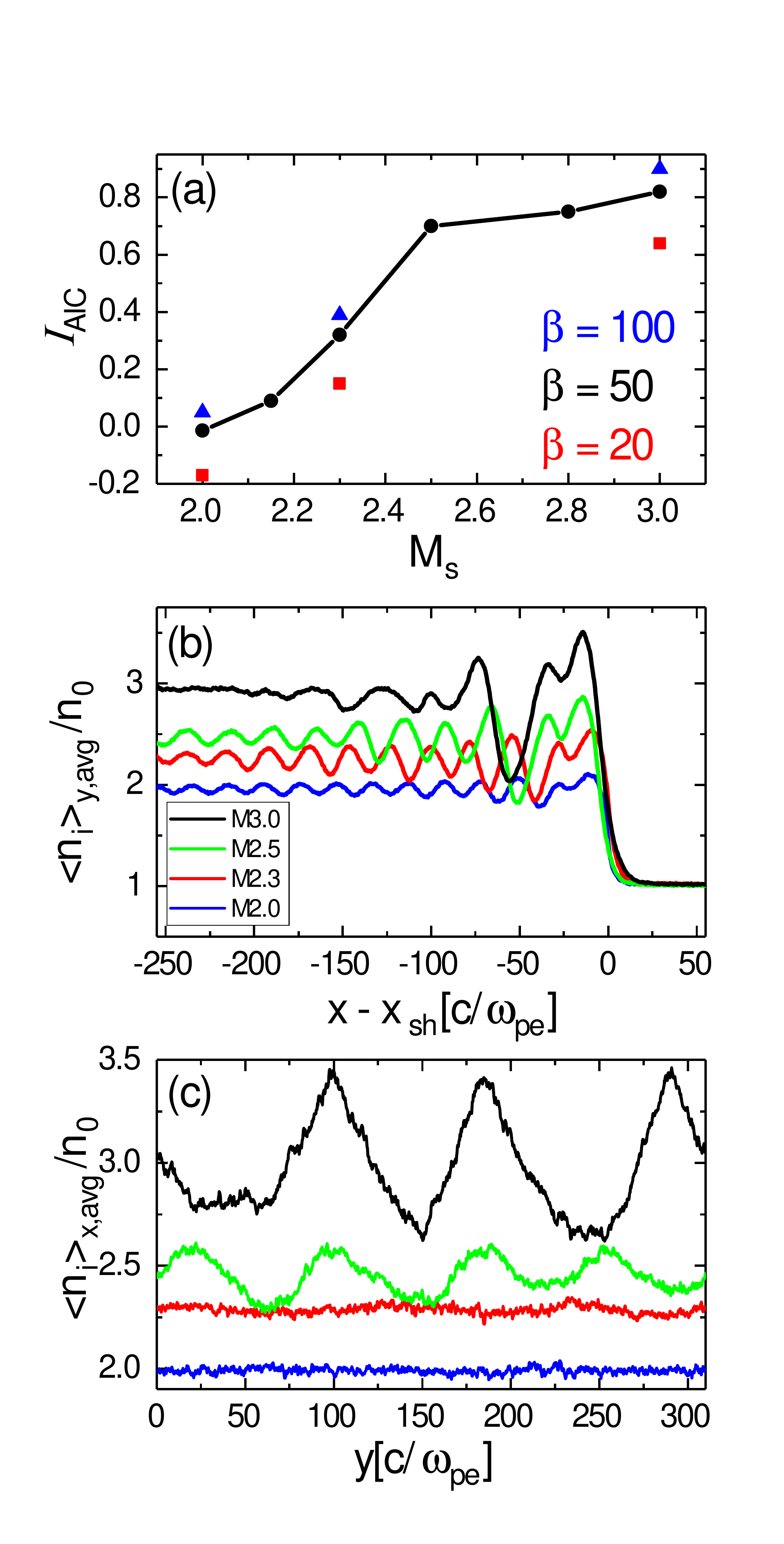}}
\vskip -0.5  cm
\caption{
(a) Instability parameter, $I_{\rm AIC}$, in Equation (\ref{eq:e3}), estimated using the mean temperature and plasma beta, 
$\langle T_{\rm i\perp}\rangle$, $\langle T_{\rm i\parallel}\rangle$ and $\langle \beta_{\rm i\parallel}\rangle$, 
in the region of $(x - x_{\rm sh})\omega_{\rm pe}/c=[-50,0]$ and $y\omega_{\rm pe}/c=[0,310]$. 
The results for the fiducial models with $\beta=50$ are shown by
the black circles connected with the black line,
while the models with $\beta=100$ and $20$ are presented by
the blue triangles and red squares, respectively.
(b) Ion number density, $\langle n_{\rm i}\rangle_{y,{\rm avg}}(x)$, averaged over $ y\omega_{\rm pe}/c=[0,310]$ in the fiducial models, M2.0(blue), M2.3 (red), M2.5 (green), and M3.0 (black). 
(c) Ion number density, $\langle n_{\rm i}\rangle_{x,{\rm avg}}(y)$, averaged over $(x - x_{\rm sh})\omega_{\rm pe}/c=[-50,0]$ in the same fiducial models as in panel (b). 
For all the quantities, the simulation results at $\Omega_{\rm ci}t \sim 32$ are used.\label{fig:f3}}
\end{figure}

Using both linear theory and hybrid simulations, \citet{gary1997} presented the instability condition for
the AIC instability: 
\begin{equation}
\label{eq:e3}
I_{\rm AIC} = \frac{T_{\rm i \perp}}{T_{\rm i \parallel}} - 1 - \frac{S_p}{\beta_{\rm  i\parallel}^{\alpha_p}} > 0,
\end{equation}
where $\beta_{\rm i\parallel} = 8\pi n_{\rm i} k_{\rm B}T_{\rm i\parallel}/B_{0}^2$ is the ion $\beta$ parallel to the magnetic field.
The fitting parameters are $\alpha_p\approx 0.72$ and $S_p\approx 1.6$ for $\beta_{\rm i\parallel}\approx 5-50$ (see their Figure 8).
This condition signifies that the AIC instability tends to be stabilized at lower $\beta_{\rm i\parallel}$ due to the stronger magnetization of ions.
For a given value of $\beta_{\rm i\parallel}$, the AIC growth rate increases with increasing $\mathcal{A}_i$, 
which in turn depends on the fraction of reflected ions.
Sine the ion temperature anisotropy is higher at stronger shocks,
the transition zone is expected to be more unstable against the AIC instability in shocks with higher $M_{\rm s}$.

Here, using the simulation results for the anisotropy $\mathcal{A}_i$, 
we calculate the instability parameter, $I_{\rm AIC}$, which is shown in Figure \ref{fig:f3}(a).
For $M_{\rm s} = 2$, $I_{\rm AIC} \lesssim 0$, so the AIC instability is stable,
which is consistent with the smooth shock structure shown in Figure \ref{fig:f2}. 
The instability parameter increases steeply around $M_{\rm s}\sim 2.2-2.4$.
Considering also the simulation results described in the next section, we suggest that the critical Mach number to trigger the AIC instability is 
$M_{\rm AIC}^{*} \approx 2.3$ in these high-$\beta$ shocks.
We note that this is similar to the critical Mach number for ion reflection, $M_{\rm s}^*$, reported in KRH2019, 
because the AIC instability is triggered by the shock-reflected ions.

\begin{deluxetable*}{ccccccc}[t]
\tablecaption{{\color{black}Linear Predictions}\label{tab:t2}}
\tabletypesize{\small}
\tablecolumns{10}
\tablenum{2}
\tablewidth{0pt}
\tablehead{
\colhead{Model Name} &
\colhead{$\mathcal{A}_{\rm i}$$^a$} &
\colhead{$\mathcal{A}_{\rm e}$$^a$} &
\colhead{AIC} &
\colhead{ion-mirror} &
\colhead{whistler} &
\colhead{electron-mirror} \\
\colhead{} &
\colhead{} &
\colhead{} &
\colhead{(${\gamma_m\over\Omega_{\rm ci}}$,${ck_m}\over \omega_{\rm pi}$)$^b$} &
\colhead{(${\gamma_m\over\Omega_{\rm ci}}$,${ck_m}\over \omega_{\rm pi}$)$^b$} &
\colhead{(${\gamma_m\over\Omega_{\rm ce}}$,${ck_m}\over \omega_{\rm pe}$)$^b$} &
\colhead{(${\gamma_m\over\Omega_{\rm ce}}$,${ck_m}\over \omega_{\rm pe}$)$^b$}
}
\startdata
LM2.0$\beta50$-m50 & 1.2 & 1.1  & stable & (0.021,0.24) & (0.030, 0.30) & (0.0066,0.21)\\
LM2.0$\beta100$-m50  & 1.2 & 1.1    & quasi-stable  & (0.034,0.21) & (0.080,0.30) & (0.018,0.23)\\
\hline
LM2.3$\beta20$-m50  & 1.5 & 1.1   & (0.10,0.34)   & (0.10,0.45) & (0.0040,0.42) & stable \\
LM2.3$\beta50$-m50   & 1.5 & 1.2   & (0.12,0.24)   & (0.14,0.34) & (0.077,0.38) & (0.019,0.29) \\
LM2.3$\beta100$-m50 & 1.5 & 1.2   & (0.14,0.18)   & (0.16,0.26) & (0.14, 0.38) & (0.041,0.27)\\
\hline
LM3.0$\beta20$-m50  & 2.0 & 1.2    &  (0.38,0.50)   &  (0.28,0.55) & (0.047,0.52) & (0.0048,0.29)\\
LM3.0$\beta50$-m50  & 2.0 & 1.2 & (0.44,0.35)  & (0.33,0.40) & (0.18,0.50) & (0.045,0.38)\\
LM3.0$\beta100$-m50 & 2.0 & 1.2  &  (0.47,0.26)  &  (0.36,0.31) & (0.30,0.50) & (0.09,0.36)\\
\hline
LM3.0$\beta50$-m100 & 2.0 & 1.2 & (0.44,0.35) & (0.33,0.40) &(0.20,0.50) & (0.048,0.38)\\
LM3.0$\beta50$-m1836 & 2.0 & 1.2 & (0.44,0.35) & (0.36,0.42) &(0.22,0.50) & (0.048,0.38)\\
\hline
LM3.0$\beta50$-m50-$\theta$53$^c$ & 1.8 & 1.1 & (0.36,0.35) & (0.27,0.38) &(0.018,0.38) & (0.006,0.24)\\
LM3.0$\beta50$-m50-$\theta$73$^c$ & 2.1 & 1.5 & (0.39,0.31) & (0.36,0.42) &(0.59,0.69) & (0.19,0.50)
\enddata
{\color{black}
\tablenotetext{a}{$\mathcal{A}_a \equiv T_{a \perp}/T_{a \parallel}$ is estimated using the simulation data for the corresponding shock model, except for LM3.0$\beta50$-m1836. The values for LM3.0$\beta50$-m100 are adopted for LM3.0$\beta50$-m1836.}
\tablenotetext{b}{Here, the values are normalized with the upstream frequencies. They are converted from the values in Table 2 of KHRK2021, in which the downstream
frequencies (e.g., $\Omega_{\rm ci,2}$ and $\omega_{\rm pe,2}$) were used. 
In KHRK2021, the spatially averaged quantities in the transition zone were given as $n_{i,2}=n_{e,2}\approx r n_0$ and $B_2\approx r B_0$. 
Thus, for example, $\Omega_{\rm ci,2}\approx r \Omega_{\rm ci}$ and $\omega_{\rm pe,2}\approx \sqrt{r} \omega_{\rm pe}$.}
\tablenotetext{c}{These two models are added in this paper.}}
\vspace{-0.8cm}
\end{deluxetable*}

As can be inferred from Equation (\ref{eq:e3}), 
Figure \ref{fig:f3}(a) shows that with similar anisotropy $\mathcal{A}_i$'s, 
$I_{\rm AIC}$ decreases as $\beta$ decreases from 100 (blue triangles) to 20 (red squares),
owing to stronger magnetization at lower $\beta$.
Thus, we expect that the AIC critical Mach number would be somewhat higher at lower-$\beta$ shocks. 
For example, \citet{hellinger1997} estimated $M_{\rm AIC}^{*}\sim 4$ for $\beta \approx 1$, using hybrid simulations, and
\citet{trotta2019} obtained a similar value, $M_{\rm AIC}^{*}\sim 3.5$ also for $\beta \approx 1$, as mentioned before.
 
Figure \ref{fig:f3}(b) shows the $y$-averaged ion number density profile, $\langle n_{\rm i}\rangle_{y,{\rm avg}}(x)$, for the models with $M_{\rm s}=2-3$.
It demonstrates that the shock becomes supercritical for $M_{\rm s} \gtrsim 2.3$,
developing substantial overshoot/undershoot oscillations in the shock transition.  
Figure \ref{fig:f3}(c), on the other hand, shows the ion number density profile, $\langle n_{\rm i}\rangle_{x,{\rm avg}}(y)$, averaged over the shock transition zone in the $x$-direction, along the $y$ axis (parallel to the shock surface) for the same four models.
The mean wavelengths of the shock surface ripples are 
$\lambda \sim 100 c/\omega_{\rm pe}\sim r_{\rm L,i}$ for the M3.0 model (black),
while $\lambda \sim 75 c/\omega_{\rm pe}\sim r_{\rm L,i}$ for the M2.5 model (green).
Note that here the Larmor radius of incoming ions, $r_{\rm L,i}\propto u_0 \propto M_{\rm s}$, scales approximately with the shock Mach number.
The variation of $\langle n_{\rm i}\rangle_{x,{\rm avg}}$ along the shock surface is insignificant for $M_{\rm s} \lesssim 2.3$.

\begin{figure*}[t]
\vskip -0.2 cm
\hskip -0.2cm
\centerline{\includegraphics[width=0.99\textwidth]{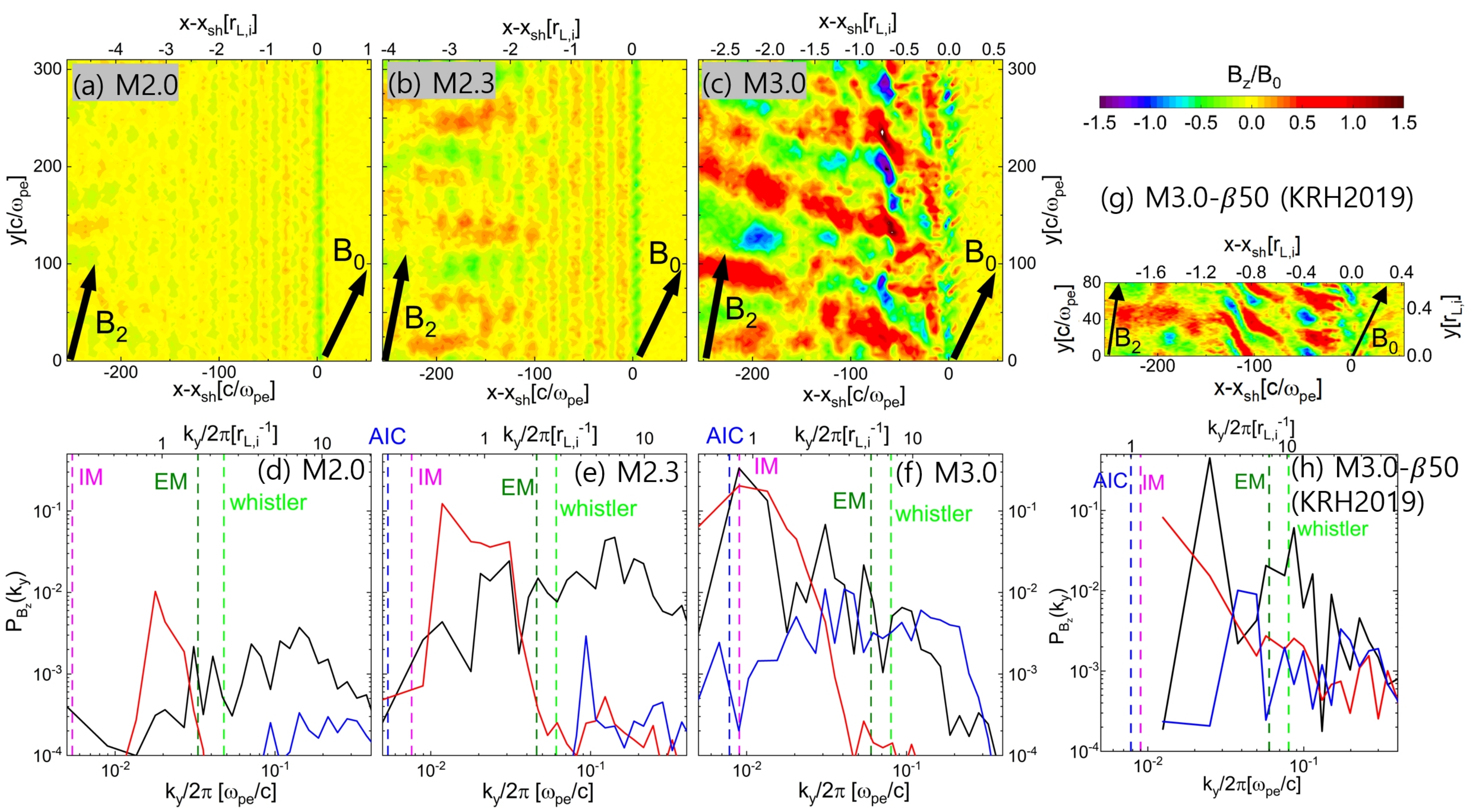}}
\vskip -0.0 cm
\caption{
Panels (a)$-$(c): Magnetic field fluctuations, $B_{z}/B_0$, in the region of $(x - x_{\rm sh})\omega_{\rm pe}/c =[-255,+55]$ at $\Omega_{\rm ci}t \sim 32$ for the fiducial models with $M_{\rm s}=2$, 2.3, and 3.
{\color{black}
The black arrows indicate the directions of the preshock and postshock magnetic field vectors, ${\bf{B_0}}$ and ${\bf{B_2}}$,
respectively.} 
Panels (d)$-$(f): Magnetic power spectrum, $P_{B_z}(k_{y}) \propto (k_{y}/2\pi) (\delta B_{z}(k_{y})^2/B_0^2)$, calculated for the shock transition region of 
$(x -x_{\rm sh})/r_{\rm L,i} \approx [-1.0,0.0]$ (black), the far downstream region of $(x -x_{\rm sh})/r_{\rm L,i} \approx [-2.8,-1.8]$ (red), and {\color{black}the upstream region of 
$(x -x_{\rm sh})/r_{\rm L,i} \approx [0.2,1.2]$ (blue)} at $\Omega_{\rm ci}t \sim 32$. 
{\color{black}The blue, bright green, magenta, and dark green vertical lines denote the wavenumbers of the maximum growth rates for the AIC, whistler, i-mirror, and e-mirror instabilities, respectively. 
Panels (g)$-$(h): Magnetic field fluctuations and power spectrum of the M3.0 model in KRH2019.}
Here, $r_{\rm L,i} \approx 91 (c/\omega_{\rm pe})\cdot (M_{\rm s}/3) \sqrt{\beta/50}\sqrt{(m_i/m_e)/50}$.
\label{fig:f4}}
\end{figure*}

\subsection{Plasma Waves in Shock Transition}
\label{sec:s4.2}

As discussed in Section \ref{sec:s2.3}, in the shock transition region of supercritical shocks,
the ion temperature anisotropy ($\mathcal{A}_i$) can trigger the AIC and ion-mirror instabilities, 
while the electron temperature anisotropy ($\mathcal{A}_e$) can induce the whistler and electron-mirror instabilities.
Due to the large mass ratio, typically electron-driven waves grow much faster on much smaller scales, 
compared to ion-driven waves.

{\color{black}
We here briefly review the linear analysis presented in KHRK2021, since the linear predictions could 
provide a useful hint on the most dominant modes as well as the early development of relevant instabilities.
Table 2 lists the anisotropies, $\mathcal{A}_{\rm i}$ and $\mathcal{A}_{\rm e}$, estimated using the shock simulation data, 
the maximum normalized growth rates, $\gamma_m/\Omega_{\rm ci}$ ($\gamma_m/\Omega_{\rm ce}$), 
and the corresponding normalized wavenumbers, $ck_m /\omega_{\rm pi}$ ($ck_m /\omega_{\rm pe}$), 
of the two ion-driven (electron-driven) instabilities.
For the fiducial M3.0 model, for instance,
the growth rates of the four instabilities have the following order:}
\begin{equation}
\label{eq:e4}
\gamma_{\rm WI} \gg \gamma_{\rm EM} \gg \gamma_{\rm AIC} > \gamma_{\rm IM}.
\end{equation}
The growth time scales of the instabilities, $\tau_{\rm inst} \equiv 1/\gamma_{\rm inst}$, 
expressed in units of $\Omega_{\rm ci}^{-1}$, are $\tau_{\rm WI}\approx 0.11$,  $\tau_{\rm EM}\approx 0.44$,
$\tau_{\rm AIC}\approx 2.3$, and $\tau_{\rm IM}\approx 3.0$ for the M3.0 model.
Hence, the whistler waves appear earlier along the first overshoot immediately downstream of the shock ramp,
while the AIC-driven waves develop much later throughout the shock transition region as ripples along the shock surface.
{\color{black}For the fiducial M2.3 model, the two ion-driven modes have comparable growth rates, which are smaller than those for the M3.0 model. 
For the fiducial M2.0 model, by contrast, the AIC instability is stable, while the growth rates of the other three instabilities are relatively low.
Thus, in subcritical shocks with $M_s\approx2$, electron-scale waves are expected to be excited mainly by the whistler instability,
while ion-scale fluctuations could be induced somewhat weakly by the ion-mirror instability.}

{\color{black}
The magnetic field fluctuations excited by the parallel AIC mode and the 
oblique ion-mirror mode can be visualized with the transverse component, $B_{z}/B_{0}$, and 
the compressional component, $(B_y-B_2)/B_0$, respectively, 
where $B_2$ is the strength of the downstream magnetic field \citep[e.g.][]{hellinger1997,guo2017}.
We find that in the M3.0 model, the wave power of the $B_z$ component is higher by an order of magnitude than that of the $B_y$ component.
Thus, we here focus mainly on the $B_z$ component excited by the AIC instability.
Panels (a)$-$(c) of Figure \ref{fig:f4} show $B_{z}/B_{0}$ for the three fiducial models.}
In the M3.0 model, we observe multi-scale waves with the wavelengths ranging from electron to ion scales 
in the shock transition region, $(x -x_{\rm sh})/r_{\rm L,i} \approx [-1, 0]$. 
As indicated by Equation (\ref{eq:e4}), in this supercritical shock,
we expect that ion-scale fluctuations are induced dominantly by the AIC instability,
while electron-scale waves are induced mainly by the whistler instability.
In Figure \ref{fig:f4}(c), we can identify such AIC-driven waves propagating mainly in the direction parallel to ${\bf{B_0}}$.
{\color{black}For comparison, the M3.0$\beta$50 model ($\beta=50$, $m_i/m_e=100$) of KRH2019 is shown in panel (g),
which exhibits features only on small scales due to the limited computation box size.
In the M2.3 model in panel (b), on the other hand, the AIC mode is not observed, whereas oblique waves driven by the ion-mirror instability appear in the far downstream region ($x -x_{\rm sh}<-2.0r_{\rm L,i}$).}
In the M2.0 model in panel (a), 
primarily electron-scale waves with small amplitudes are observed in the shock transition,
while weak oblique waves probably due to the ion-mirror instability appear in the far downstream region.

The black lines in Figure \ref{fig:f4}(d)$-$(f) show the magnetic power spectra, $P_{\rm B_{z}}$, 
in the shock transition region shown in the upper panels (a)(c). 
In the M3.0 model, $P_{\rm B_{z}}$ indicates the presence of multi-scale waves in the wide range of
wavenumbers, $k_{y}/2\pi \sim [0.009 - 0.9] \omega_{\rm pe}/c$, 
corresponding to $\lambda \sim [11.1 - 111] c/\omega_{\rm pe}$.
In particular, the wavelength of the ion-scale waves driven by the AIC instability, $\lambda \sim 111 c/\omega_{\rm pe} \sim 1.2 r_{\rm L,i}$, is similar to the size of shock surface surface ripples, $\lambda_{\rm ripple}$, as shown in Figure \ref{fig:f2}. 
The spectrum also shows the substantial powers of the electron-scale waves with $k_{y}/2\pi \sim 0.09 \omega_{\rm pe}/c$ ($\lambda \sim 11.1 c/\omega_{\rm pe}$) driven by the whistler instability.
{\color{black}Again, for comparison, $P_{\rm B_{z}}$ for the M3.0$\beta$50 model of KRH2019 is shown in panel (h).
Due to the small transverse domain, the powers on small wavenumbers, $k_{y}/2\pi\lesssim 10^{-2} \omega_{\rm pe}/c$, are not present.}
In the M2.3 model in panel (e), the electron-scale waves are relatively more important than the ion-scale waves,
while the ion-scale waves are driven mainly by the ion-mirror instability as indicated in panel (b).
In the M2.0 model in panel (d), the ion-scale waves are almost absent. 
These results are consistent with the fact that multi-scale plasma waves can be triggered by the AIC instability only in supercritical shocks with $M_{\rm s} \gtrsim 2.3$.

The plasma waves induced in the shock transition undergo nonlinear evolution, while being advected further downstream.
In order to investigate such nonlinear evolution, 
in Figure \ref{fig:f4}(d)$-$(f), we also show $P_{\rm B_{z}}$ (red lines) in the far downstream region of 
$(x - x_{\rm sh})/r_{\rm L,i} \approx [-2.8, -1.8]$. 
A few points are noted: (1) The whistler waves are excited in the transition region immediately behind 
the ramp, and then undergo rapid damping via electron scattering, 
leading to the reduction of $\mathcal{A}_e$ in the far downstream region.
So the magnetic power of electron-scale waves is significantly reduced there as well. 
(2) In the M3.0 model in panel (f), $P_{\rm B_{z}}$ of ion-scale waves remains relatively substantial in the far downstream, even after experiencing nonlinear evolution. 
(3) In the M2.0 and M2.3 models in panels (d) and (e), $P_{\rm B_{z}}$ on small $k$'s is mainly due to the oblique waves excited by the ion-mirror instability in the far downstream region,
which can be seen in panels (a) and (b).

{\color{black}In the case of the $M_{\rm s}=3$ shock with $\theta_{\rm Bn}=73^{\circ}$ in \citet{niemiec2019}, 
the SDA reflection of electrons is ineffective during the early stage, since the de Hoffman–Teller velocity is greater than the electron thermal velocity, 
i.e., $u_{\rm t} \equiv u_{\rm s}/\cos \theta_{\rm Bn} \sim 1.5 v_{\rm th,e}$ \citep[see also][]{guo2014b}. 
After $t\gtrsim \tau_{\rm AIC}$, the electron reflection is enhanced (see their Figure 2) due to locally reduced magnetic fields and/or locally reduced $\theta_{\rm Bn}$,
caused by the rippled shock surface.
By contrast, in our $M_{\rm s}=3$ shock model with $\theta_{\rm Bn}=63^{\circ}$, $u_{\rm t}\sim v_{\rm th,e}$ and the fraction of reflection electrons is substantial ($\sim23$\%) even 
before the emergence of the shock ripples (see Fig. 1 of KRH2019).}
{\color{black}In fact, the enhancement of the electron reflection fraction and the ensuing EFI-driven waves in the upstream region due to the shock ripples is only a few \%  (also compare the blue lines in panels (f) and (h) of Figure \ref{fig:f4}).
Although the fraction of suprathermal electrons is increased slightly by the addition of SSDA 
in the new simulations (see Figure \ref{fig:f6}), the impacts on the self-excited upstream waves seem only marginal, possibly 
due to the limited integration time.}

\begin{figure*}[t]
\vskip -0.0 cm
\hskip -0.2 cm
\centerline{\includegraphics[width=0.95\textwidth]{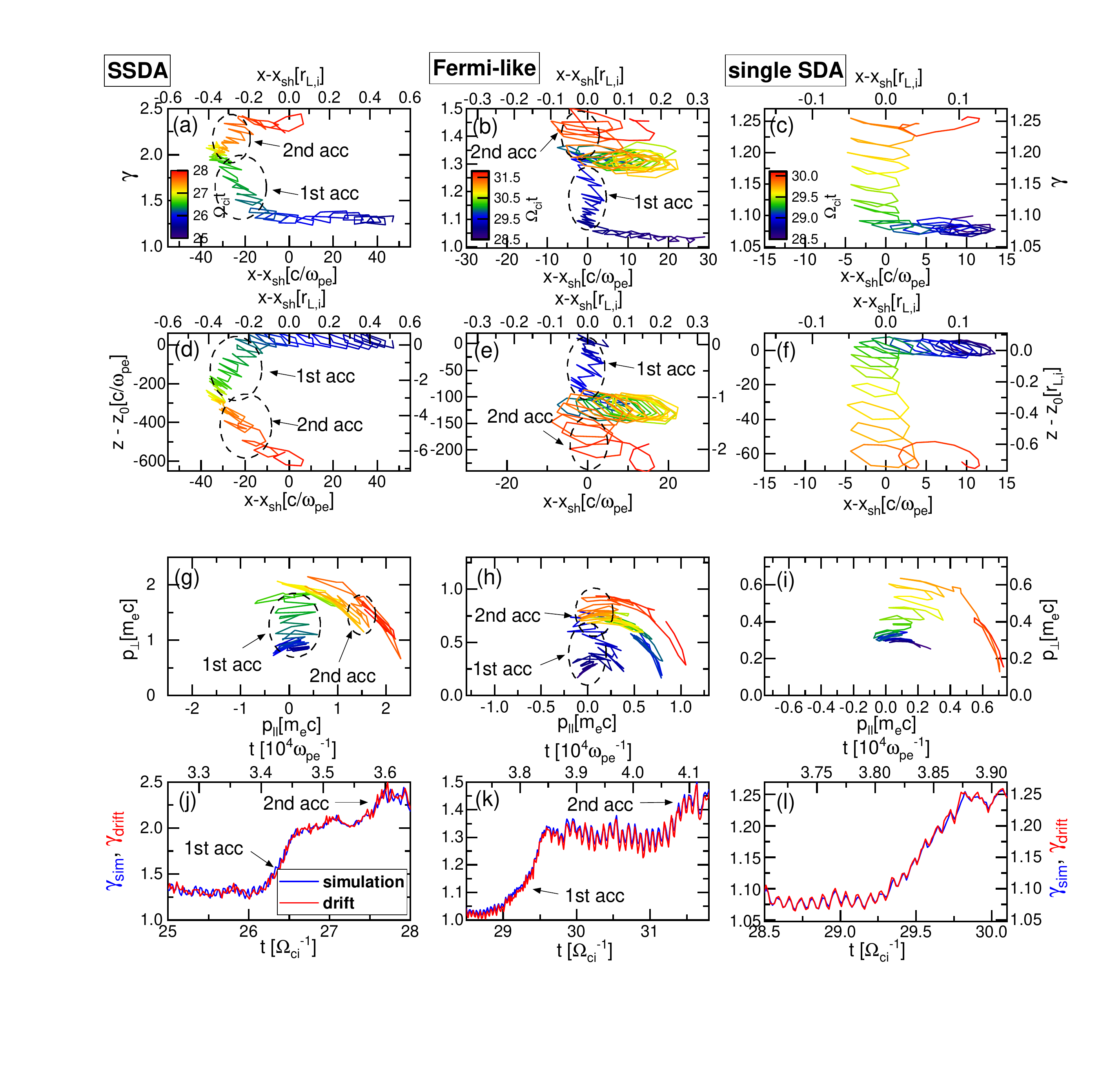}}
\vskip -1.5 cm
\caption{{\color{black}Left panels: Trajectory of a selected electron that undergoes the SSDA
during $\Omega_{\rm ci}t \sim 25 - 29.5$. 
Middle panels: Trajectory of a selected electron that undergoes the Fermi-like acceleration
during $\Omega_{\rm ci}t \sim 28 - 32$.
Right Panels: Trajectory of a selected electron that undergoes a single cycle of the SDA
during $\Omega_{\rm ci}t \sim 28.5 - 30.1$.
{\color{black}Note that the trajectories in the shock rest frame are shown.}
They are taken from the M3.0 model simulation.
In panels (g)$-$(i), the trajectories along the $p_{\perp}$-direction show the energy gain due to the gradient-B drift along the
motional electric field, while the trajectories along the arcs in the $p_{\parallel}-p_{\perp}$ space represent pitch-angle scattering.
In panels (j)$-$(l), the blue lines show the evolution of the Lorentz factor, $\gamma_{\rm sim}$, in the simulation frame,
while the red lines show the energy gain, $\gamma_{\rm drift} = -(e/m_{\rm e}c^2)\int E_{z}dz$, 
estimated using the motional electric field in the shock transition zone.\label{fig:f5}}}
\end{figure*}

\begin{figure}[t]
\vskip -0.0 cm
\hskip -0.2 cm
\centerline{\includegraphics[width=0.5\textwidth]{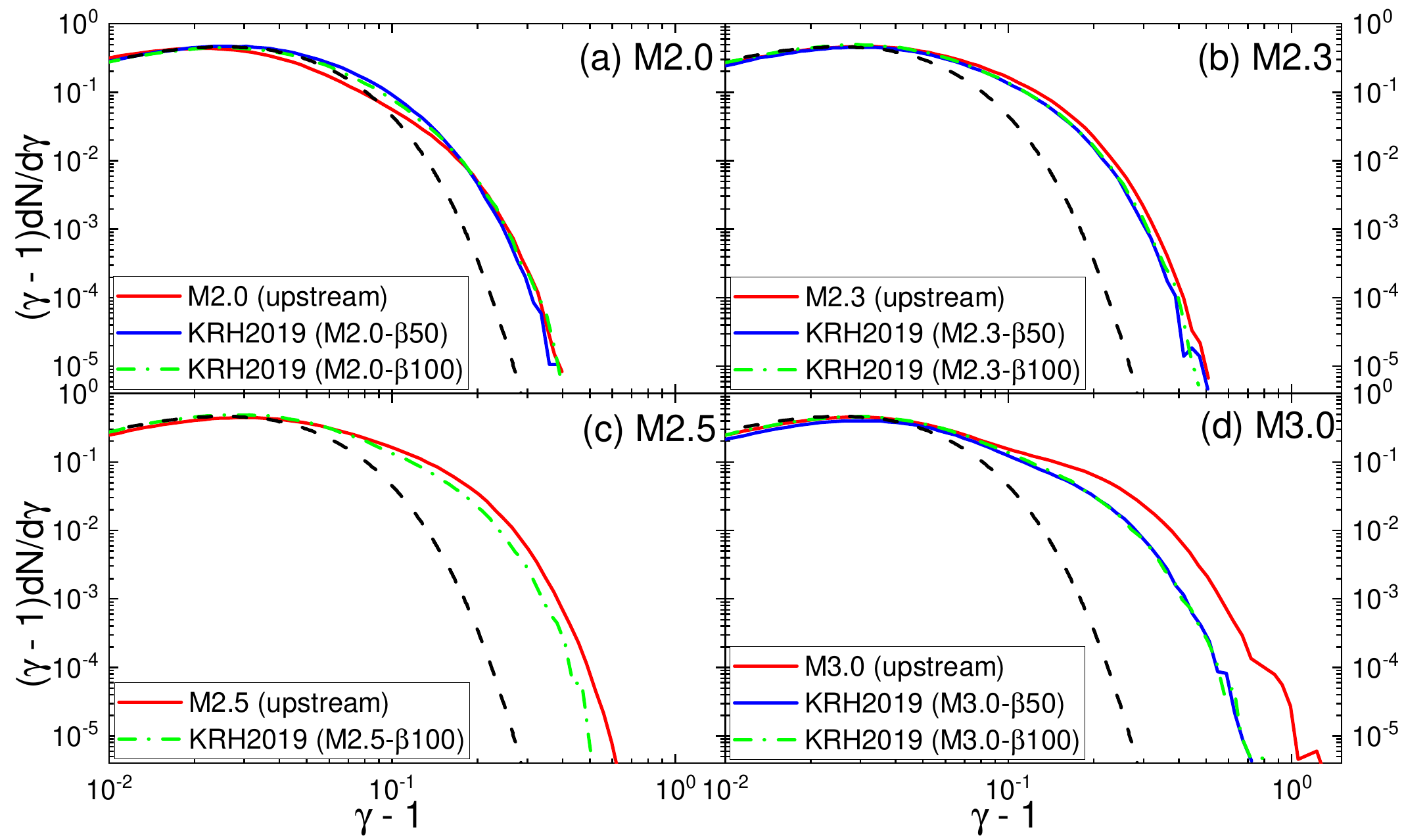}}
\vskip -0.2 cm
\caption{Upstream electron energy spectra (red solid lines) at $\Omega_{\rm ci}t \sim 32$ in the
fiducial models with $M_{\rm s}=2-3$. 
The spectra are taken from the region of $(x-x_{\rm sh})/ r_{\rm L,i}=[0,+1]$ and the black dashed lines show the Maxwellian distributions in the upstream. 
The blue solid ($\beta=50$) and green dot-dashed ($\beta=100$) lines show the upstream electron energy spectra 
at $\Omega_{\rm ci}t \sim 30$ for the models in KRH2019,
in which the transverse domain, $L_y/r_{\rm L,i}$, is about 8 times smaller than that of the simulations in this study. \label{fig:f6}}
\end{figure}

\subsection{Electron Preacceleration via SSDA}
\label{sec:s4.3}

To understand the preacceleration mechanism in our simulations, we examine how electrons gain energy in the M3.0 model shock.
{\color{black}
Figure \ref{fig:f5} shows the trajectories of three selected electrons that gain energy via the SSDA (left panels), the Fermi-like acceleration (middle panels), and the SDA (right panels).
{\color{black}Note that the trajectories in the shock rest frame are shown; the region of 
$(x - x_{\rm sh})\omega_{\rm pe}/c \approx [-5,5]$ corresponds to the shock ramp, 
while the downstream region of $(x - x_{\rm sh})\omega_{\rm pe}/c \approx [-50,0]$ contains 
both the first and second overshoots in the transition zone (see Figure 2(b)).
The trajectory in panels (a) and (d) shows that this electron is confined within the transition zone during $\Omega_{\rm ci}t\sim 26 - 28$,
and undergoes the first (green) and second (orange) stages of the gradient-B drift along the $-z$-direction,
illustrating the SSDA.
Panels (b) and (e) show that the electron experiences the Fermi-like acceleration by going 
through the first (blue) and second (mainly orange to reddish orange) stages of the SDA, 
while being reflected at the ramp and scattered by upstream waves.
Panels (c) and (f) show that the electron undergoes only a single cycle of the standard SDA.}
In panels (g)$-$(i), the trajectories roughly parallel to the $p_{\perp}$-direction show the energy gain due to the drift
along the motional electric field, while the trajectories following the arcs in the $p_{\parallel}-p_{\perp}$ 
space represent pitch-angle scattering.
In panels (g) and (h), the electrons experience two episodes of acceleration, as indicated by the ellipses and arrows.}

The bottom panels, (j)$-$(l), compare the variation of the Lorentz factor in the simulation (blue line) with the
energy gain of $\gamma_{\rm drift} = -(e/m_{\rm e}c^2)\int E_{z}dz$ (red line), which is
expected to accumulate from the drift along the motional electric field in the shock transition zone.
We confirm that the preacceleration realized in the simulated shock is consistent 
with the SSDA mechanism proposed by previous studies \citep{katou2019,niemiec2019}. 
Although electrons can be energized by both the Fermi-like acceleration and SSDA,
the most energetic electrons are produced mainly by the SSDA. 
 
Figure \ref{fig:f6} compares the upstream electron energy spectra in the fiducial models of the current study (red lines) 
{\color{black}with the corresponding spectra for the models with $\beta=50$ (blue lines) and $\beta=100$ (green lines) 
reported in KRH2019.
Note that $m_i/m_e=100$ in KRH2019, but the simulations were insensitive to the mass ratio \citep[see also][]{guo2014b}.}
As mentioned before, in KRH2019, the 2D simulation domain was too small in the transverse direction 
to include the emergence of the shock surface rippling via the AIC instability.
As a result, the SDA-reflected electrons gain energy only through the Fermi-like acceleration in that study.
The figure clearly demonstrates that in the case of supercritical shocks with $M_s=2.5$ and 3, 
through the SSDA, electrons can be accelerated further to higher energies in the new simulations than in the simulations of KRH2019. 
In subcritical shocks, on the other hand, the AIC instability is not triggered and the ensuing SSDA does not occur
even in the new simulations with a larger simulation domain.

The PIC simulation of a $\beta\approx 5$ shock by \citet{niemiec2019} and the hybrid
simulations of $\beta\approx1$ shocks by \citet{trotta2019} showed that electrons could be preaccelerated
well above the injection momentum through the SSDA in supercritical shocks.
{\color{black}
In \citet{niemiec2019}, for instance, the highest electron energy of $\gamma_{\rm max}\approx 60$ was achieved by
several phases of the SSDA for the simulation duration of $\Omega_{\rm ci}t_{\rm max} \approx 79$, 
while $\gamma_{\rm inj}\approx 25$ for their shock parameters.
By contrast, Figure \ref{fig:f5} shows only two phases of the SSDA for our M3.0 model, resulting in $\gamma_{\rm max}\sim 2.5$.} 
So the electron energy spectrum even in the new simulation (the red of Figure \ref{fig:f6}(d)) is extended 
to the energy below the injection momentum ($\gamma_{\rm inj}\sim 7$).
This should be the limitation of the PIC simulation for the model.
In fact, simulating electron energization all the way to injection to DSA in $\beta\approx 100$ shocks
would require much larger simulation domains and much longer simulation times.
{\color{black}However, we expect that the SSDA would continue to reach $p> p_{\rm inj}$, as long as electrons are confined 
within the shock transition region by scattering due to multi-scale waves.}

\subsection{Dependence on the Model Parameters}
\label{sec:s4.4}

\begin{figure*}[t]
\vskip -0.2 cm
\hskip -0.2 cm
\centerline{\includegraphics[width=0.8\textwidth]{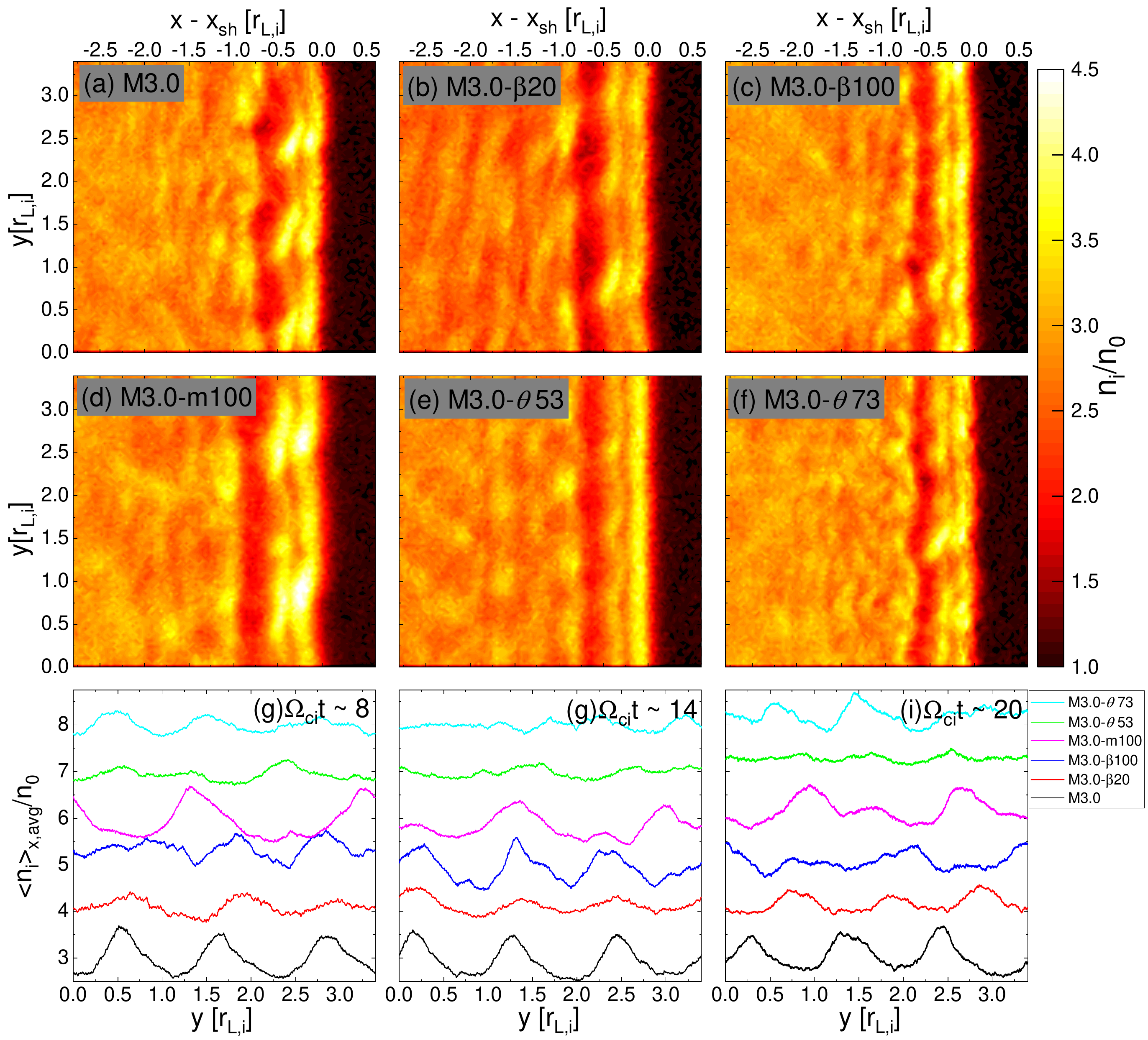}}
\vskip -0.0 cm
\caption{Panels (a)$-$(f): Ion number density, $n_{\rm i}(x,y)/n_0$, in the region of 
$(x - x_{\rm sh})/r_{\rm L,i}=[-2.8,0.6]$ and $y/r_{\rm L,i}=[0,3.4]$ at $\Omega_{\rm ci}t \sim 20$
in the six M3.0 models with different values of $\beta$, $m_i/m_e$, and $\theta_{\rm Bn}$.
The fiducial M3.0 model has $M_{\rm s}=3$, $\beta=50$, $m_i/m_e=50$, and $\theta_{\rm Bn}=63^{\circ}$.
See Table 1 for the parameters of other models. 
{\color{black}Panels (g)$-$(i): Ion number density, $\langle n_{\rm i}\rangle_{x,{\rm avg}}(y)$, averaged over $(x - x_{\rm sh})/r_{\rm L,i}=[-0.6,0]$ at $\Omega_{\rm ci}t \sim 8-20$ for the same set of the models.} 
{\color{black}The line color for each model is given in the small box.
Each line is shifted vertically by $+1$ for the purpose of clarity. 
Note that the growth time scale for the AIC instability is $\Omega_{\rm ci}\tau_{\rm AIC}= 2.1-2.6$, 
so the density distribution shown here displays the fluctuations in nonlinear stages.}
\label{fig:f7}}
\end{figure*}

In this section, we examine how our findings depend on the simulation parameters, such as, $\beta$, $m_i/m_e$, and $\theta_{\rm Bn}$.
{\color{black}Table \ref{tab:t2} summarizes the parameter dependence of the linear predictions for the ion-driven and electron-driven instabilities.
For all the four instabilities, $\gamma_m/\Omega_{\rm ci}$ or $\gamma_m/\Omega_{\rm ce}$ are higher and $ck_m /\omega_{\rm pi}$ or $ck_m /\omega_{\rm pe}$ are smaller for higher $\beta$.
Regarding the dependence on the mass ratio, 
both $\gamma_m/\Omega_{\rm ci}$ and $ck_m /\omega_{\rm pi}$ are almost independent of $m_i/m_e$ for the AIC mode, 
whereas they are slightly higher for larger $m_i/m_e$ for the ion-mirror mode.
For the whistler and electron-mirror modes, $\gamma_m/\Omega_{\rm ce}$ is slightly higher for larger $m_i/m_e$, while $ck_m /\omega_{\rm pe}$ is almost independent of $m_i/m_e$ (see also Figure 3 of KHRK2021).}

{\color{black}
The superluminal obliquity angle is $\theta_{\rm sl}\equiv \arccos(u_{\rm sh}/c)\approx 84^{\circ}$ for the M3.0 model,
hence the electron reflection fraction rate, $R$, multiplied by the mean SDA energy gain, $\langle \Delta \gamma \rangle$, is expected to increases
as $\theta_{\rm Bn}$ increases from $53^{\circ}$ to $73^{\circ}$ (see Figure 1(c) of KRH2019).}
{\color{black}So the temperature anisotropies, $\mathcal{A}_{\rm i}$ and $\mathcal{A}_{\rm e}$, 
are slightly higher for larger $\theta_{\rm Bn}$.
The linear predictions in Table \ref{tab:t2} show that both $\gamma_m$ and $k_m$ increase with increasing
$\theta_{\rm Bn}$, except for the AIC mode.
Especially, the growth rate increases drastically with increasing $\theta_{\rm Bn}$ for the electron-driven modes.
However, the enhanced electron anisotropy suppresses the AIC instability, so $\gamma_{\rm AIC}$ 
is the greatest for $\theta_{\rm Bn}=63^{\circ}$.}

{\color{black}
We point that the comparison of the PIC simulations with the linear predictions should be made
with the following caveats.
In the linear analysis, the background plasma is assumed to be spatially homogeneous and electrons and ions are prescribed with bi-Maxwellian velocity distributions.
In the shock transition zone, on the other hand, the density and magnetic field distributions are non-uniform with the overshoot/undershoot oscillations and
the velocity distributions of ions and electrons are non-Maxwellian due to the reflected particles that are
convected downstream.
Moreover, even weak $Q_{\perp}$-shocks usually exhibit time-varying behaviors.}

\begin{figure*}[t]
\vskip -0.2 cm
\hskip -0.2cm
\centerline{\includegraphics[width=0.95\textwidth]{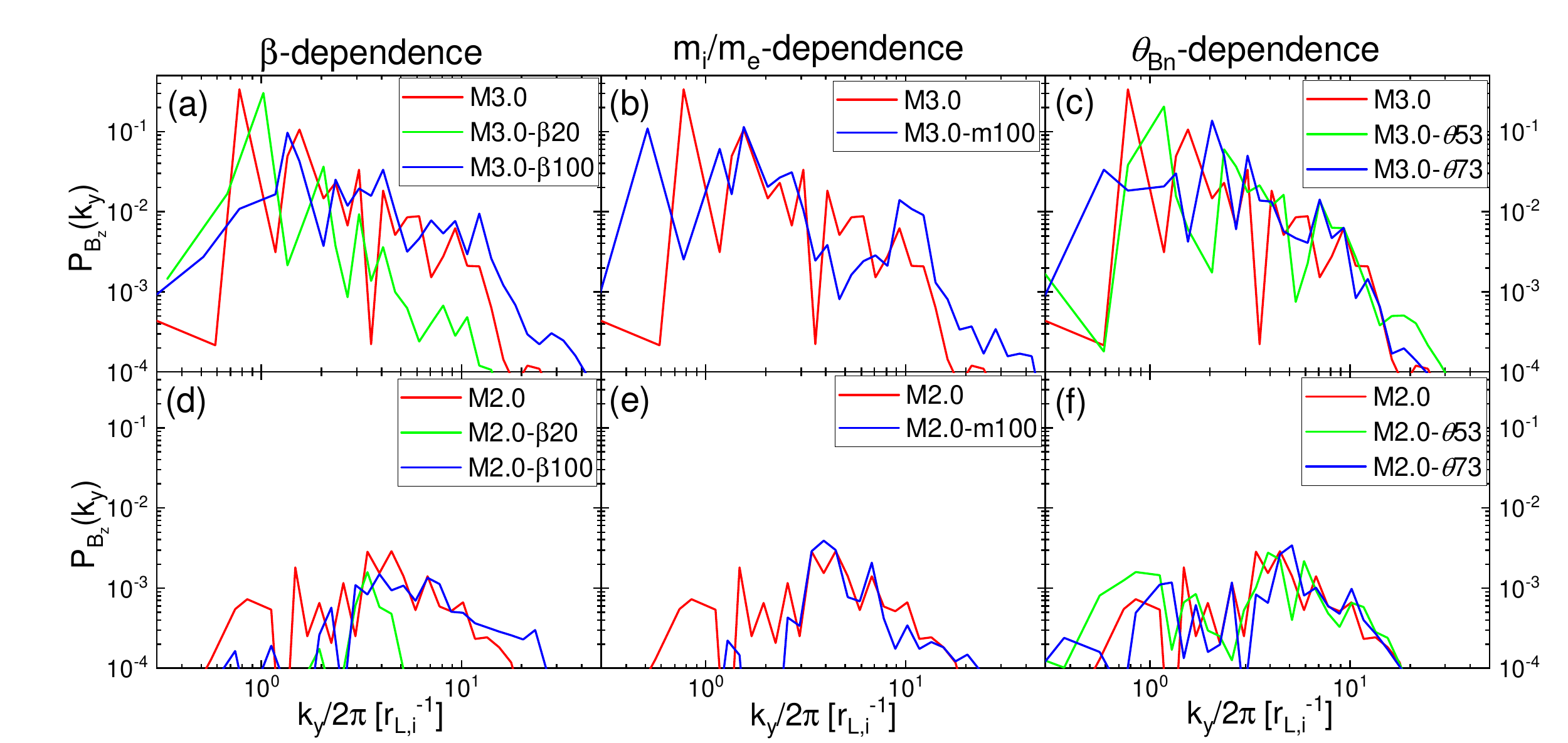}}
\vskip -0.0 cm
\caption{Magnetic power spectra, $P_{B_z}(k_{y}) \propto (k_{y}/2\pi) (\delta B_{z}(k_{y})^2/B_0^2)$, at $\Omega_{\rm ci}t \sim 20$
for the M3.0 models (upper panels) and the M2.0 models (lower panels) with different parameters,
in the transition region of $(x - x_{\rm sh})/r_{\rm L,i}=[-1.0,0.0]$. 
In the fiducial M2.0 and M3.0 models, $\beta=50$, $m_i/m_e=50$, and $\theta_{\rm Bn}=63^{\circ}$.
See Table 1 for the parameters of other models.
Note that here the wavenumber $k_{y}$ is normalized with the Larmor radius for incoming ions, 
$r_{\rm L,i} \approx 91 (c/\omega_{\rm pe})\cdot (M_{\rm s}/3)\sqrt{\beta/50}\sqrt{(m_i/m_e)/50}$.
\label{fig:f8}}
\end{figure*}

In Figure \ref{fig:f7}, we compare the ion density distribution in the six M3.0 models with 
different parameters.
{\color{black}Here, the length is normalized with $r_{\rm L,i}\propto \sqrt{\beta}\sqrt{(m_i/m_e)}$. }
{\color{black}
Panels (g)$-$(i) show  the ion number density, $\langle n_{\rm i}\rangle_{x,{\rm avg}}$, averaged over $(x - x_{\rm sh})/r_{\rm L,i}=[-0.6,0]$ at $\Omega_{\rm ci}t \sim 8-20$.
We note that this time period is much longer than
the growth time scale of the AIC instability, $\Omega_{\rm ci}\tau_{\rm AIC}\approx 2.1-2.6$.
Hence, panels (g)$-$(i) illustrate the time-varying density configuration in fully developed, nonlinear stages of the AIC instability.
According to the linear predictions given in Table \ref{tab:t2}, 
$\lambda_{\rm AIC}/r_{\rm L,i}\propto (\lambda_{\rm AIC}\omega_{\rm pi}/c)/\sqrt{\beta}$ is independent 
of $m_i/m_e$, but decreases only slightly with $\beta$, i.e., $\lambda_{\rm AIC}/r_{\rm L,i}\sim 1.5$, 1.4, and 1.3 for $\beta=20$, 50, and 100, respectively.
The wavelengths of induced ripples, $\lambda_{\rm ripple} \sim 1.0 - 1.7 r_{\rm L,i}$ agree reasonably well with the linear analysis. 
As mentioned in Section \ref{sec:s4.1}, we expect that, due to the periodic boundary condition,
about $2-3$ ripples would appear along the $y$-direction. For example,
in the M3.0 model (black lines) the dominant mode has $\lambda \sim L_y/3$, 
whereas the modes with $\lambda \sim L_y/2$ and $L_y/3$ look dominant in the M3.0-m100 model (magenta lines).

Regarding the dependence on $\beta$, 
the M3.0 model with $\beta=50$ (black) shows somewhat larger fluctuations than the two models with $\beta=20$ (red) and $100$ (blue),
although $\gamma_{\rm AIC}/\Omega_{\rm ci}$ increases with increasing $\beta$.
So the amplitudes of the dominant modes do not seem to match exactly the linear predictions.
Considering that $\gamma_{\rm AIC}/\Omega_{\rm ci}$ in the $\beta=100$ model is higher by about 7\% 
than that of the $\beta=50$ model, this discrepancy might be due to the smoothing effects of the enhanced electron-driven modes 
at high $\beta$, as well as possible nonlinear effects.
Regarding the dependence on $\theta_{\rm Bn}$, the M3.0 model with $\theta_{\rm Bn}=63^{\circ}$ (black) shows greater amplitudes than the two models with $\theta_{\rm Bn}=53^{\circ}$ (green) and $73^{\circ}$ (cyan),
which is consistent with the behavior of $\gamma_{\rm AIC}/\Omega_{\rm ci}$ in Table \ref{tab:t2}.}

Figure \ref{fig:f8} compares the power spectra of magnetic field fluctuations for the M3.0 and M2.0 models with different values of $\beta$, $m_i/m_e$ and $\theta_{\rm Bn}$. 
In all the models with $M_{\rm s}=3$ (upper panels), multi-scale waves in the range of 
$k_{y}r_{\rm L,i} /2\pi \sim [0.7 - 10]$
($\lambda \sim [0.1 - 1.5] r_{\rm L,i}$) are present.
{\color{black}In particular, in the fiducial M3.0 model with $\beta=50$ and $\theta_{\rm Bn}=63^{\circ}$, $P_{\rm B_z}$ looks somewhat larger on the AIC-driven scales than in other models, which is consistent with the visual impression of the density fluctuations shown in Figure \ref{fig:f7}.}
In all the models with $M_{\rm s}=2$ (lower panels), by contrast,
mainly electron-scale waves are excited, as expected. 

\begin{figure*}[t]
\vskip -0.2 cm
\hskip -0.3 cm
\centerline{\includegraphics[width=1.08\textwidth]{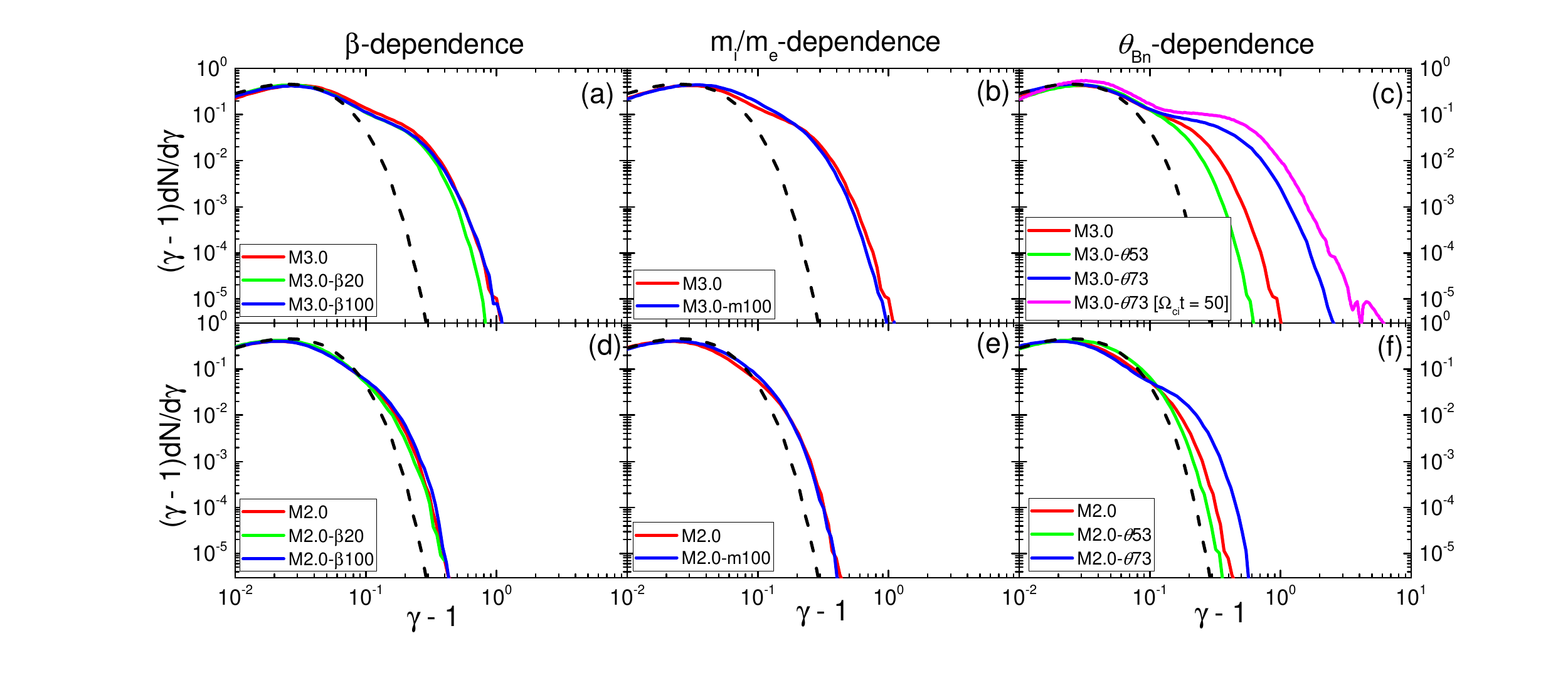}}
\vskip -0.5 cm
\caption{Upstream electron energy spectra at $\Omega_{\rm ci}t \sim 20$ 
for the M3.0 models (upper panels) and the M2.0 models (lower panels) with different parameters.
The spectra are taken from the region of $(x-x_{\rm sh})/ r_{\rm L,i}=[0,+1]$ and the black dashed lines show the Maxwellian distributions in the upstream. 
In the fiducial M2.0 and M3.0 models, $\beta=50$, $m_i/m_e=50$, and $\theta_{\rm Bn}=63^{\circ}$.
See Table 1 for the parameters of other models.
In panel (c), the magenta line shows the energy spectrum at $\Omega_{\rm ci}t \sim 50$ for the M3.0-$\theta$73 model.
\label{fig:f9}}
\end{figure*}

In Figure \ref{fig:f9}, we examine the upstream electron energy spectra for the same set of the models shown in Figure \ref{fig:f8}.
The figure shows that the preacceleration depends only weakly on $\beta$ and 
$m_{\rm i}/m_{\rm e}$, while it is more efficient with larger $\theta_{\rm Bn}$.
{\color{black}This is mainly because the motional electric field is stronger for higher $\theta_{\rm Bn}$, so the SDA-reflected electrons are more energetic.} 
In the M3.0-$\theta$73 model, in which the simulation was carried out for a longer time, $\Omega_{\rm ci}t\sim 50$ (magenta line in Figure \ref{fig:f9}(c)),
some of the most energetic electrons were accelerated to $p_{\rm inj} \sim 3p_{\rm th,i}$ ($\gamma_{\rm inj} \sim 7$).
This implies that the preacceleration via the SSDA could be a feasible mechanism for electron injection to the full DSA process
in high-$\beta$ supercritical shocks,
as previously shown for lower $\beta$ shocks in \citet{niemiec2019} and \citet{trotta2019}. 
In all the models with $M_{\rm s}=2$, however, the energy spectra seem consistent with the single SDA 
cycle \citep[e.g.,][]{guo2014b}, and neither the Fermi-like acceleration nor the SSDA is effective.

{\color{black}Figure \ref{fig:f9} demonstrates that the preacceleration of electrons and the shock criticality are 
almost independent of $m_i/m_e$,
but depend somewhat weakly on $\beta~(\approx 20-100)$ for the ranges of values considered here. }
Furthermore, the preacceleration would be more effective at larger $\theta_{\rm Bn}$, as long as 
the shock parameters satisfy the subluminal condition, i.e., $\theta_{\rm Bn} \le \arccos(u_{\rm sh}/c)$
(see KRH2019).

\section{Summary}
\label{sec:s5}

In supercritical $Q_{\perp}$-shocks, a substantial fraction of incoming ions and electrons are 
reflected at the shock ramp \citep[e.g.][]{krasnoselskikh2013}. 
The gyromotion of the reflected ions in the immediate upstream and downstream of the shock ramp generates the foot and the transition 
structures, respectively (see Figure \ref{fig:f1}).
The reflected electrons backstreaming along the background magnetic field can experience the Fermi-like acceleration in the shock foot \citep{guo2014a,guo2014b,kang2019},
whereas the downstream advected electrons may undergo the SSDA in the shock transition  \citep{katou2019,niemiec2019}.
In both the acceleration mechanisms, the primary energy source is the gradient drift along the motional electric field in the shock transition.
In the Fermi-like acceleration, electrons are scattered back and forth between the shock ramp and the upstream waves self-generated 
via the EFI in {\color{black}the shock upstream region}.
In the SSDA, electrons undergo stochastic pitch-angle scattering off the multi-scale waves, 
which are excited by both the ion and electron temperature anisotropies, $T_{\rm i\perp}/T_{\rm i\parallel}$ and $T_{\rm e\perp}/T_{\rm e\parallel}$,
in the shock transition.

In KRH2019, we performed 2D PIC simulations to study electron preacceleration 
in $Q_{\perp}$ ICM shocks.
We found that the Fermi-like acceleration involving multiple SDA cycles 
can operate only in supercritical shocks with the sonic Mach number greater than the critical Mach number 
$M_{\rm ef}^*\approx 2.3$.
In this work, with the specific aim to examine the SSDA, aided by ion-driven instabilities,
we extended the work of KRH2019 by considering the 2D simulation domain large enough to properly encompass
ion-scale waves in the transverse direction.
As a result, the new set of simulations can include the excitation of the multi-scale waves from the electron skin depth to the ion Larmor radius, 
$c/\omega_{\rm ep} \lesssim \lambda \lesssim r_{\rm L,i}$.

The main results can be summarized as follows:\hfill\break
1. Adopting the numerical values for $T_{\rm i \perp}$, $T_{\rm i \parallel}$, and $\beta_{\rm i\parallel}$
in the shock transition zone of the simulated models with $M_{\rm s}=2-3$,
we estimate the instability parameter, $I_{\rm AIC}$, defined in Equation (\ref{eq:e3}).
Considering both the behavior of $I_{\rm AIC}$ and the PIC simulation results, 
we suggest that the critical Mach number, above which the AIC mode becomes unstable, is $M_{\rm AIC}^*\approx 2.3$
for $\beta~\approx 20-100$.
Note that in this study, the critical Mach number is defined in terms of the sonic Mach number rather than the
Alfv\'enic Mach number,
since both ion and electron reflections are controlled mainly by the shock compression.\hfill\break
2. The simulations confirm
that overshoot/under-shoot oscillations and shock surface rippling become increasingly 
more evident for higher $M_{\rm s}$ in supercritical shocks with $M_{\rm s} \gtrsim M_{\rm AIC}^*$, 
while the shock structure seems relatively smooth for subcritical shocks.
\hfill\break
3. In the transition zone of supercritical shocks, 
ion-scale waves can be generated by the AIC and ion-mirror instabilities due to the ion temperature anisotropy ($T_{\rm i\perp}/T_{\rm i\parallel} > 1$), while 
electron-scale waves can be generated by the whistler and electron-mirror instabilities due to the electron temperature anisotropy ($T_{\rm e\perp}/T_{\rm e\parallel} > 1$).
Both the linear analysis and the periodic-box PIC simulations presented in KHRK2021 indicate that
the AIC and whistler instabilities are dominant over the ion and electron mirror instabilities, respectively,
in high-$\beta$ plasmas under consideration.
In the case of subcritical shocks with small anisotropies, on the other hand, primarily electron-scale waves are 
induced by the whistler instability, while ion-scale waves with small amplitudes could be excited by the ion-mirror instability.\hfill\break
4. In $\beta\approx 20-100$ supercritical shocks, electrons are confined within the shock transition for an extended period
and gain energy by the SSDA, as suggested by previous studies for $\beta\sim 1-5$ shocks
\citep{niemiec2019,trotta2019}. 
Although we could not see electron preacceleration all the way
to injection to DSA in our PIC simulations due to the limited simulation domain and time,
we suggest that the combination of the Fermi-like acceleration and the SSDA could
energize thermal electrons to the full DSA regime in supercritical, $Q_{\perp}$-shocks in the ICM.  \hfill\break
5. The shock criticality in terms of triggering the AIC instability (or shock surface rippling) 
depends rather weakly on $m_i/m_e$ and $\theta_{\rm Bn}$
for the ranges of values considered here.
However, the critical Mach number, $M_{\rm AIC}^*$, tends to be somewhat higher at lower $\beta$ ($\sim 1$) owing to the
stronger magnetization of ions \citep[e.g.,][]{trotta2019}.
In addition, the preacceleration of electrons is relatively insensitive to $\beta$ ($\sim 20-100$) and $m_i/m_e$,
while its efficiency increases with increasing $\theta_{\rm Bn}$, as long as the shock remains subluminal.
Therefore, we infer that our findings about the shock criticality and the preacceleration 
can be applied generally to $Q_{\perp}$-shocks in the ICM. 
\hfill\break

Finally, radio and X-ray observations indicate that some radio relics
seem to have $M_{\rm s} \sim 1.5-2.3$ \citep[e.g.,][]{vanweeren2019}.
Then, an outstanding problem is how to explain the origin of such radio relics.
We suggest that further studies should be done to explore how additional ingredients, such as preexisting fossil CR electrons and/or 
preexisting turbulence on kinetic plasma scales, could influence electron acceleration in subcritical ICM shocks. 
Furthermore, it would be interesting to consider a scenario, 
in which electrons, preaccelerated in locally supercritical shocks,
are injected to DSA at subcritical portions,
since the simulations of the formation of large-scale structures including galaxy clusters indicate that the shock surfaces associated with radio relics could be highly nonuniform
with varying $M_{\rm s}$ and $\theta_{\rm Bn}$ \citep[e.g.,][]{hong2015,roh2019}. 

\acknowledgments
{\color{black}The authors thank the anonymous referee for constructive comments.}
This research used the high performance computing resources of the UNIST Supercomputing Center. J.-H.H. and D.R. were supported by the National Research Foundation (NRF) of Korea through grants 2016R1A5A1013277 and 2020R1A2C2102800. J.-H. H. was also supported by the Global PhD Fellowship of the NRF through grant 2017H1A2A1042370. 
S.K. was supported by the NRF grant funded by the Korea government (MSIT) (NRF-2020R1C1C1012112).
H.K. was supported by the Basic Science Research Program of the NRF through grant 2020R1F1A1048189.

\bibliography{ms_shock}{}
\bibliographystyle{aasjournal}

\end{document}